\documentclass{nature}

\usepackage{graphicx}%
\usepackage{multirow}%
\usepackage{amsmath,amssymb,amsfonts}%
\usepackage{amsthm}%
\usepackage{mathrsfs}%
\usepackage[title]{appendix}%
\usepackage{xspace}
\usepackage{xcolor}%
\usepackage{textcomp}%
\usepackage{manyfoot}%
\usepackage{booktabs}%
\usepackage{algorithm}%
\usepackage{algorithmicx}%
\usepackage{algpseudocode}%
\usepackage{listings}%
\usepackage{rotating}
\usepackage[strings]{underscore}
\usepackage{tabularx}
\usepackage[strict]{changepage}  
\usepackage{fontawesome} 
\usepackage{pifont}  
\usepackage{tcolorbox}  
\usepackage{tikz} 
\usepackage[normalem]{ulem} 
\usepackage{newtxtext,newtxmath} 
\usepackage{xurl}  
\usepackage{subcaption}

\definecolor{wyf}{RGB}{214, 160, 29} 
\definecolor{wyfnew}{RGB}{36, 116, 181} 
\definecolor{newlyadded}{HTML}{000000} 
\definecolor{agentbackground}{RGB}{224, 241, 248} 
\definecolor{agenttext}{RGB}{32, 109, 142}
\definecolor{toolbackground}{HTML}{26428B} 
\definecolor{tooltext}{HTML}{26428B} 
\definecolor{codebackground}{RGB}{249, 249, 249} 
\definecolor{codetext}{HTML}{014f86} 
\definecolor{actionbackground}{RGB}{255, 250, 205} 
\definecolor{actiontext}{HTML}{312E81}
\definecolor{number}{HTML}{D3E1AB} 
\definecolor{character}{HTML}{C2B8E0} 

\newcommand{\newlyadded}[1]{\textcolor{newlyadded}{#1}} 
 
\newcommand{\wyf}[1]{\textcolor{black}{#1}}

\newcommand{\ie}{i.e.}
\newcommand{\eg}{e.g.}
\newcommand{\textsquarenumber}[2][number]{%
  \tikz[baseline=(char.base)]\node[shape=rectangle, fill=#1, text=black, inner sep=2pt, font=\footnotesize\bfseries\ttfamily] (char) {\textbf{\texttt{#2}}};}
  \newcommand{\textsquarecharacter}[2][character]{%
  \tikz[baseline=(char.base)]\node[shape=rectangle, fill=#1, text=black, inner sep=2pt, font=\footnotesize\bfseries\ttfamily] (char) {\textbf{\texttt{#2}}};}
\newcommand{\systemname}{\textit{Nexus}\xspace}

\usepackage{tcolorbox}
\tcbuselibrary{skins}

\newcommand{\planner}{Planner\xspace}
\newcommand{\executor}{Executor\xspace}
\newcommand{\evaluator}{Evaluator\xspace}

\newcommand{\toolname}[1]{#1\xspace}
\newcommand{\code}[1]{\textit{#1}\xspace}

\newcommand{\actionname}[1]{\textit{#1}\xspace}

\usepackage{lineno}

\usepackage{setspace}
\usepackage{amsfonts}
\usepackage{amssymb}
\usepackage{amsmath}

\usepackage{float}
\usepackage{placeins}
\usepackage{graphicx}

\usepackage[export]{adjustbox}

\usepackage[
    font=normalfont,
    labelfont=bf,
    labelformat=empty
]{caption}

\newcounter{mybodyfigure}
\newcounter{myedfigure}

\newcommand{\beginbodyfigures}{\renewcommand{\thefigure}{{\themybodyfigure}}}

\newcommand{\beginedfigures}{\renewcommand{\thefigure}{{\themyedfigure}}}

\newcommand{\stepbodyfigure}{\refstepcounter{mybodyfigure}}

\DeclareRobustCommand{\bodyfigure}[1]{%
  \stepbodyfigure%
  \label{#1}%
  \themybodyfigure%
}

\newcommand{\bodyfigurelabel}[1]{\textbf{Figure \bodyfigure{#1}:}}

\makeatletter
\g@addto@macro\caption@prepareslc{%
  \renewcommand{\stepbodyfigure}{\caption@l@stepcounter{mybodyfigure}}}
\makeatother

\newcommand{\stepedfigure}{\refstepcounter{myedfigure}}

\makeatletter
\g@addto@macro\caption@prepareslc{%
  \renewcommand{\stepedfigure}{\caption@l@stepcounter{myedfigure}}}
\makeatother

\usepackage[super,comma,sort&compress]{natbib}

\usepackage{hyperref}
\usepackage{verbatim}

\usepackage{ulem} 
\usepackage{rotating} 
\usepackage{soul}

\usepackage{color} 

\usepackage[colorinlistoftodos]{todonotes}

\makeatletter
\newsavebox\myboxA
\newsavebox\myboxB
\newlength\mylenA

\newcommand*\xoverline[2][0.75]{%
    \sbox{\myboxA}{$\m@th#2$}%
    \setbox\myboxB\null
    \ht\myboxB=\ht\myboxA%
    \dp\myboxB=\dp\myboxA%
    \wd\myboxB=#1\wd\myboxA
    \sbox\myboxB{$\m@th\overline{\copy\myboxB}$}
    \setlength\mylenA{\the\wd\myboxA}
    \addtolength\mylenA{-\the\wd\myboxB}%
    \ifdim\wd\myboxB<\wd\myboxA%
       \rlap{\hskip 0.5\mylenA\usebox\myboxB}{\usebox\myboxA}%
    \else
        \hskip -0.5\mylenA\rlap{\usebox\myboxA}{\hskip 0.5\mylenA\usebox\myboxB}%
    \fi}
\makeatother

\title{Figures as Interfaces: Toward LLM-Native Artifacts for Scientific Discovery}

\author{Yifang Wang$^{1,2,3,4,5}$, Rui Sheng$^{6}$, Erzhuo Shao$^{1,3,4,7}$, Yifan Qian$^{1,3,4,5}$, Haotian Li$^{6}$, Nan Cao$^{8}$, and Dashun Wang$^{1,3,4,5,7*}$}

\begin{document}

\maketitle

{\small
\begin{affiliations}
 \item Center for Science of Science and Innovation, Northwestern University, Evanston, IL, USA
 \item Department of Computer Science, Florida State University, Tallahassee, FL, USA
 \item Northwestern Innovation Institute, Northwestern University, Evanston, IL, USA
 \item Ryan Institute on Complexity, Northwestern University, Evanston, IL, USA
 \item Kellogg School of Management, Northwestern University, Evanston, IL, USA
 \item Department of Computer Science and Engineering, Hong Kong University of Science and Technology, Hong Kong, China
 \item McCormick School of Engineering, Northwestern University, Evanston, IL, USA
 \item Intelligent Big Data Visualization Lab, Tongji University, Shanghai, China
 
 *Correspondence to: \href{dashun.wang@kellogg.northwestern.edu}{dashun.wang@kellogg.northwestern.edu}
\end{affiliations}}


\begin{abstract}
Large language models (LLMs) are transforming scientific workflows, not only through their generative capabilities but also through their emerging ability to use tools, reason about data, and coordinate complex analytical tasks. 
Yet in most human-AI collaborations, the primary outputs, figures, are still treated as static visual summaries: once rendered, they are handled by both humans and multimodal LLMs as images to be re-interpreted from pixels or captions. 
The emergent capabilities of LLMs open an opportunity to fundamentally rethink this paradigm. 
In this paper, we introduce the concept of LLM-native figures: data-driven artifacts that are simultaneously human-legible and machine-addressable. 
Unlike traditional plots, each artifact embeds complete provenance: the data subset, analytical operations and code, and visualization specification used to generate it. 
As a result, an LLM can ``see through'' the figure---tracing selections back to their sources, generating code to extend analyses, and orchestrating new visualizations through natural-language instructions or direct manipulation. 
We implement this concept through a hybrid language–visual interface that integrates LLM agents with a bidirectional mapping between figures and underlying data. 
Using the science of science domain as a testbed, we demonstrate that LLM-native figures can accelerate discovery, improve reproducibility, and make reasoning transparent across agents and users. 
More broadly, this work establishes a general framework for embedding provenance, interactivity, and explainability into the artifacts of modern research, redefining the figure not as an end product, but as an interface for discovery.
For more details, please refer to the demo video available at \href{https://llm-native-figure.com/}{www.llm-native-figure.com}.
\end{abstract}

\setcounter{mybodyfigure}{0}
\beginbodyfigures


\section{Introduction}
\label{section:introduction}

Figures play a central role in scientific discovery~\cite{anscombe1973graphs}. 
They summarize complex data, reveal patterns, and communicate results in a visual form that is both concise and interpretable~\cite{anscombe1973graphs}. 
A figure is often the endpoint of analysis and the centerpiece of scientific reasoning that transforms data into insight. 
\wyf{Yet for all their importance, figures remain static visual summaries and are designed primarily for human interpretation.} 
They capture relationships between variables but conceal the underlying data, code, and transformations that produced them.

The emergence of large language models (LLMs) offers an opportunity to rethink this paradigm. 
Beyond text generation, LLMs increasingly serve as computational agents capable of reasoning, coding, and executing analytic workflows~\cite{openai2025deepresearch, google2025deepresearch, microsoftdiscovery, shao2025sciscigpt, tu2025towards, harvey-ai}.  
\wyf{While most current uses of LLMs in science focus on accelerating established scientific workflows~\cite{shao2025sciscigpt, lu2026towards}, a complementary question is whether the artifacts of research themselves, particularly the scientific figures, should be redesigned to internalize the capabilities these models introduce.} 
When an LLM agent generates a figure, it possesses complete knowledge of the artifact's provenance~\cite{lasser2020creating, konkol2020publishing, ziemann2023five, curvenote, e-life, pasquier2017practical, rupprecht2020improving}, including its dataset, analytical process, and visualization. 
From the model's perspective, a figure is not a static image but a structured object whose components can be queried, revised, and extended. 
This capability implies that, in principle, figures can be more interpretable to LLMs than to humans, providing a foundation for new modes of scientific reasoning and interaction.

To explore this opportunity, we introduce a framework for constructing \textit{LLM-native figures}: \wyf{research artifacts} that are simultaneously human-legible and machine-interpretable. 
\wyf{By “native,” we mean these artifacts are designed from the ground up to exploit LLM capabilities as their core computational engine, rather than merely being augmented by them.
Unlike conventional systems that treat LLMs as a convenience layer, an LLM-native figure relies fundamentally on the model to maintain a continuous, bidirectional mapping between natural-language intent, analytical operations, and visual interactions, seamlessly translating user instructions into analytical actions and decoding visual interactions directly back into executable operations.
As a result, the figure becomes a queryable, extensible, and reproducible analytical object that encapsulates its visualization, provenance, data, and executable code in a single cohesive loop. 
In addition, by linking each figure to preceding figures in the iterative analytical process, the framework enables the creation of \textit{data-driven artifacts}, composable units that record the trajectory of iterative exploration across one or more linked figures, preserving the non-linear logic of scientific discovery.}
To instantiate this framework, we develop \systemname, a proof-of-concept system for the science of science domain that implements LLM-native figures and data-driven artifacts. 
Users can interact with figures and data insights using both natural language and direct manipulations~\cite{shneiderman1983direct, ware2019information, keim2008visual, munzner2014visualization}. 
The system interprets user intent, analyzes data, and records every analytic step, including user instructions and interactions, code execution, and visualization generation and coordination, in an artifact that preserves version history and enables full reproducibility. 
We evaluate \systemname through four case studies demonstrating its ability to support multi-step scientific analysis; a computational evaluation assessing the fidelity of its bidirectional mapping; and a user study with twelve science of science (SciSci) experts examining the tool's usefulness and usability.
Together, these contributions establish a foundation for LLM-native figures that bridge human and machine reasoning. 
By transforming figures from static endpoints into interactive, provenance-aware interfaces, this work advances a general method for human-AI collaborative discovery in the era of large language models.
\section{Related Work}
\label{section:related-work}
Our framework builds on recent advances in AI-powered scientific discovery, interactive visual data exploration, and human-LLM interaction, which together provide a critical foundation for accelerating human understanding in data-rich domains. 

\textbf{LLMs and AI agents for scientific discovery.} 
\wyf{Recent breakthroughs in LLMs and AI agents are rapidly accelerating discovery across disciplines~\cite{zheng2025automation, zhang2024comprehensivesurvey, wang2023scientific}, driving innovations from drug design~\cite{zhang2025scientific, ghafarollahi2025sciagents, zheng2025large} and algorithm development~\cite{lu2026towards} to behavioral studies and broad data science~\cite{maojun2025survey, inala2024data, hong2024data, guo2024ds, shao2025sciscigpt, wang2019human, manning2024automated}.
By leveraging advanced LLM capabilities such as planning and reasoning, and external tool and skill integration~\cite{gao2025democratizing, liang2026skillnet}, these systems aim to automate full or partial research pipelines~\cite{lu2026towards}, from hypothesis generation, experimental design, code execution~\cite{guo2024ds}, and data-insight extraction~\cite{hong2024data, huang2025biomni, sphinx2026, observable2026canvas}, to figure generation~\cite{zhu2026paperbanana, plottie2026} and manuscript preparation~\cite{lu2026towards}. 
} 
These tools range from fully autonomous command-line tools and packages~\cite{lu2026towards, hong2024data, guo2024ds} 
to more recent conversational interfaces that keep humans in the discovery loop through natural language interaction~\cite{openai2025deepresearch, google2025deepresearch, microsoftdiscovery, shao2025sciscigpt}. 
%

Despite their power, most of these systems constrain the discovery process to linear, end-to-end workflows or simple conversational exchanges. This approach contrasts sharply with the inherently iterative, non-linear nature of human scientific reasoning.
Their reliance on text-based outputs, occasionally complemented by static figures or tables, establishes a ``one-shot" interaction paradigm that limits dynamic exploration and iterative reasoning. 
Furthermore, these linear workflows make it difficult to trace analytical provenance, as they lack a well-structured record of the exploration process. 
Although new forms of computationally reproducible research (\eg, Curvenote~\cite{curvenote} and eLife's computationally reproducible article~\cite{e-life}) and studies~\cite{lasser2020creating, konkol2020publishing, ziemann2023five} have been proposed to increase transparency and reproducibility in open science by using digital-coding notebooks and interactive articles to link paper text to its underlying figures, code, and data, these systems focus on the final stage of research rather than supporting the iterative data exploration that precedes it.

\textbf{Visual data exploration.} 
\wyf{Data visualization has long been recognized as a powerful tool for scientific discovery, enabling humans to uncover hidden patterns and communicate findings that traditional statistical methods might miss~\cite{anscombe1973graphs}.
Through direct manipulations~\cite{shneiderman1983direct}, interactive dashboards provide intuitive access to complex, multidimensional data essential for exploration.}
However, traditional visual analytics systems~\cite{keim2008visual}, while powerful, are rigid, task-specific, and demand substantial manual effort and domain expertise throughout development~\cite{munzner2014visualization, wang2023innovationinsights, wang2025funding}. 
\newlyadded{Commercial business intelligence (BI) platforms, such as Tableau and Power BI~\cite{tableau, powerbi}, and recent research tools~\cite{observable2026canvas, wang2023data, wang2025data} lower this development barrier by allowing analysts to assemble dashboards from tabular data without programming. 
Yet their workflows remain centered on constructing views: 
users specify data fields, transformations, and visual encodings, and the system renders the corresponding chart or dashboard.
}


Recent advances in AI and LLMs have begun to address these limitations through automated chart generation~\cite{tableauagent, tian2024chartgpt, wang2023llm4vis, dibia2023lida}, single-visualization manipulation and refinement~\cite{wang2023data, wang2025data}, multi-view dashboard creation with automated task decomposition~\cite{zhao2024lightva, zhao2024leva, lange2025yac}, and proactive AI-assisted insight exploration~\cite{zhao2025proactiveva}.
%
These tools, however, remain fundamentally limited: they either support only single-round chart generation, require advanced programming skills for chart manipulation, generate complete visualization systems based solely on users' initial queries, or support simple chart coordination based on pre-generated data. 
%
\newlyadded{Commercial BI platforms have likewise begun to integrate natural-language input~\cite{tableauagent, powerbillm}, but also primarily as an authoring aid: a prompt is translated into a chart or report page, after which the result is edited and explored through the conventional interface. 
The LLM thus remains a convenience layer beside the figure, with its outputs confined to the platform's built-in chart types and calculation language over a pre-modeled data source; interactions with the figure are never decoded back into a subsequent analysis.} 
As a result, these approaches are not able to accommodate the dynamic, open-ended nature of scientific exploration, where questions and analytical directions evolve continuously as new insights emerge. 


\textbf{Human-LLM interaction.} 
The rise of LLMs and generative AI has highlighted the need for a new way of thinking about user interface design in the human-computer interaction community, inspiring both new interaction modes~\cite{gao2024taxonomy, shen2025prompting, he2025plan, shen2026interaction} and innovative interface designs~\cite{luera2024survey, chen2025generative}. 
Among these, generative and malleable user interfaces have gained significant attention in both industry (\eg, ChatGPT Canvas~\cite{chatgpt-canvas} and Claude Artifacts~\cite{claude-artifact}) and academia~\cite{chen2025generative, cao2025generative, suh2023sensecape}. 
These interfaces enable users to iteratively generate and refine digital artifacts, such as code, documents, and dashboards, through conversational interaction with LLMs, supporting a wide range of activities from everyday tasks~\cite{cao2025generative, suh2023sensecape} to creative ideation~\cite{you2025designmanager} and data analysis~\cite{bigquery-canvases}. 
While these systems move beyond traditional chat-based interfaces by supporting iterative artifact refinement, most are designed for assembling interfaces rather than data-driven discovery. 
Tasks such as scientific exploration demand complex analytical workflows, iterative deep dives, and fine-grained control over data insights to support scientific rigor, reproducibility, and transparency. 
Existing systems either overlook these requirements or rely on users to interact directly with low-level code and data schemas~\cite{bigquery-canvases}. 


\section{Results}
\label{section:results} 
\subsection{Conceptual Framework}
\label{subsection:conceptual-framework} 
We propose a conceptual framework that redefines scientific figures as composite computational objects, \textit{LLM-native figures}, that are human-legible, machine-interpretable, and support scientific provenance.
Each LLM-native figure encapsulates the full analytical provenance underlying its visualization, linking the rendered image, the visualization specification, the generated data insight, the data subset, and the computational process that produced it. 
\wyf{The framework is powered by an underlying multi-agent LLM that enables real-time, dynamic data exploration through both natural-language queries and interactive visualizations. 
Users can progressively explore emerging questions and insights based on existing visual discoveries---an iterative process that naturally mirrors how scientists conduct discovery across diverse analytical domains.} 

We illustrate this framework through a simple example (Supplementary Note 2.1). 
Imagine a climate researcher studying historical monthly temperature records across the U.S. 
She enters a natural-language query:
\textit{``Show Florida's average temperature by month of year (Jan–Dec), averaged across 2014–2024.''}
The system responds with a line chart (Fig.~\ref{fig:Figure1}a) in which the horizontal axis denotes calendar months and the vertical axis represents the average monthly temperature in Florida. 
This figure looks entirely familiar, resembling a standard plot that scientists from many disciplines might produce. 

The interaction becomes more interesting when the researcher begins to use the figure itself as an interface to explore emerging questions. 
For example, to learn how Florida's summer temperatures compare to those of other states, she brushes over the summer months on the Florida plot and requests:
\textit{``Which states are the hottest during these brushed months? Rank all states by average temperature (2014–2024), as a vertical bar chart with states on the x-axis.''}
The system interprets this selection interaction as a request to focus on the subset of rows corresponding to June–August (Fig.~\ref{fig:Figure1}b), aggregates the data across all states (Fig.~\ref{fig:Figure1}c), and generates a new bar chart ranking states by their average summer temperature (Fig.~\ref{fig:Figure1}e). 
If she repeats the same interaction over winter months (Fig.~\ref{fig:Figure1}f), the bar chart updates to show a different ranking (Fig.~\ref{fig:Figure1}h), making it straightforward to contrast seasonal patterns without writing new queries or code. 
Because both charts are linked through a shared underlying representation, subsequent operations, such as filtering to coastal states, splitting by decade, or changing to a different aggregation, can be requested either via natural language or through further interactions with the figures.

This example captures two key capabilities of the framework: natural-language instructions that drive analytical operations and generate visualizations, and visual interactions that map directly back to the underlying data and analytical pipeline. 
These capabilities are enabled by four design components: the representation of a figure, the bidirectional mapping between visual and analytical operations, the data-driven artifact, and the LLM engine that serves as the backbone of the figure.

\begin{figure}[!t]
    \centering
    \includegraphics[width=1\linewidth]{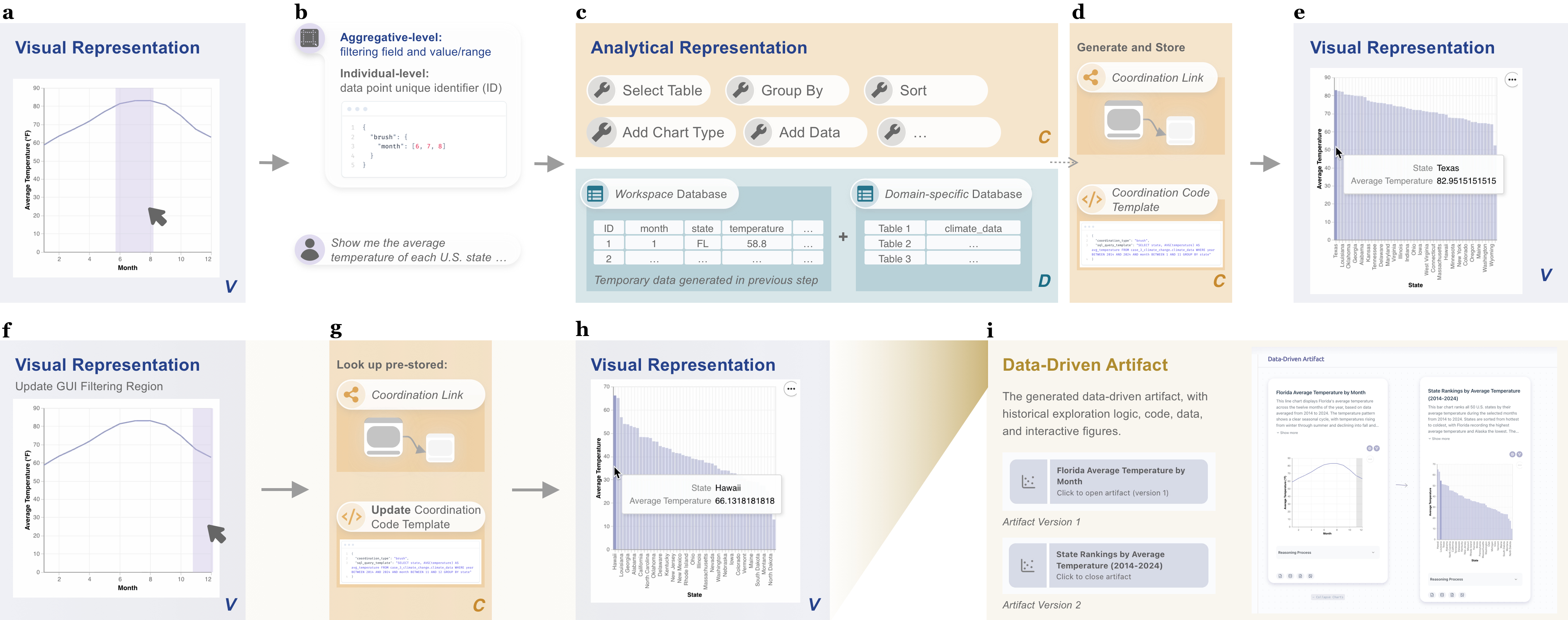}
    \vspace{1pt}
    \caption{ 
    \bodyfigurelabel{fig:Figure1}
    \textbf{Dynamic generation, coordination, and preservation of LLM-native figures.}
   \textbf{Example of iterative data exploration: }
    Starting from an existing figure that displays the 12-month temperature trend in Florida (\textbf{a}), the user brushes a temporal interval and describes the desired follow-up analysis in natural language (\textbf{b}).
    The framework uses the Visualization → Analytical operations mapping to translate these mixed-modality inputs into precise database queries (\textbf{c}),  enabling the system to retrieve the exact subset of data indicated by the user's visual and linguistic cues.  
    It then applies the Analytical operations → Visualization mapping to generate a new interactive figure that supports deeper exploration (\textbf{e}). 
   \textbf{Example of real-time coordination between linked figures: }
    When users adjust the focus in one figure (\eg, by brushing a different temporal window \textbf{(f)}), the linked figure updates automatically (\textbf{h}). 
    To enable fast, stable coordination, the system pre-computes and stores the coordination relation and executable code template at the moment the new figure is created (\textbf{d}), allowing subsequent interactions to directly retrieve and update the relevant code (\textbf{g}) to query the database.
   \textbf{Persistent, revisitable artifacts:}
    The linked figures, together with their underlying code, data, and coordination rules, are all stored in the data-driven artifact that captures the user's exploration trajectory and supports future revisits, refinements, and extensions (\textbf{i}).
    } 
\end{figure}

\textbf{Representation of LLM-Native Figures.} 
\wyf{LLM-native figures are grounded in a principle of dual legibility: they must remain human-legible through intuitive visual access while also being machine-interpretable by exposing their full analytical provenance. 
Rather than treating figures as monolithic images, we operationalize this principle by capturing them as structured analytical states that link the visual output directly to its underlying data and logic.} 
We represent each figure $F_t$ produced by the system at time $t$ as:

\begin{equation}
F_t = \{V_t, C_t, D_t, M_t\}
\end{equation}
Fig.~\ref{fig:Figure2}f illustrates this representation in the context of the example above. 
$V_t$ denotes the rendered visualization (Fig.~\ref{fig:Figure2}f-V). 
It is represented through multiple modalities: (1) the rendered raster image (\eg, PNG), (2) a textual summary of key insights, and (3) a declarative visualization specification (\eg, Vega-Lite JSON schema~\cite{satyanarayan2016vega}) \wyf{that defines how data attributes are mapped to visual channels (\eg, position, size, and color) and specifies interactive behaviors}. 
Together, these modalities make the figure not only human-readable but also computationally navigable.
Each visual mark  (\eg, point, bar, or line segment) in $V_t$ has a unique identifier that maps to corresponding rows or groups in the dataset $D_t$. 
$C_t$ records the analytical actions (Sec.~\ref{subsection:method-llm}) and corresponding executable code (\eg, SQL and Python) that generated the dataset $D_t$ and visualization $V_t$ (Fig.~\ref{fig:Figure2}f-C).
$D_t$ represents the underlying data used in the visualization, such as data subsets and associated data schema, and aggregated results used for rendering (Fig.~\ref{fig:Figure2}f-D).
$M_t$ represents the metadata associated with this figure within the user's exploration history, such as the timestamp $t$, version identifier, type and description of user operations, and links to corresponding data-driven artifacts (Fig.~\ref{fig:Figure2}f-M).
This structured representation ensures that every visual element can be traced to the data and computational steps that produced it.
\wyf{Conversely, any modification to the data or computational steps that produced it automatically updates the visualization and records the change in the metadata.} 
The tuple thus forms a closed and reproducible description of the analytical state at the moment the figure is produced.

\begin{figure}[!t]
    \centering
    \includegraphics[width=1\linewidth]{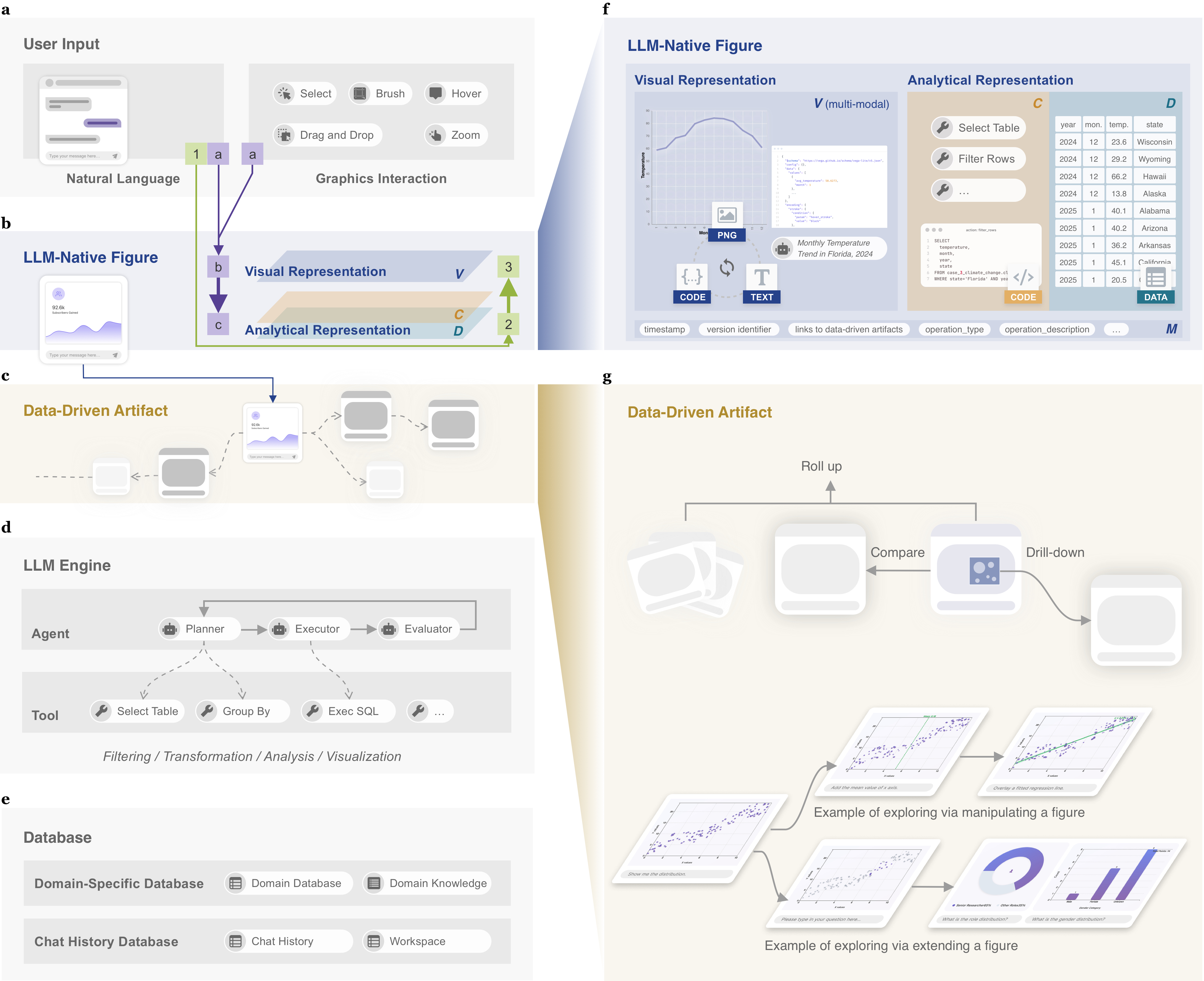}
    \vspace{1pt}
    \caption{
    \bodyfigurelabel{fig:Figure2}
   \textbf{The conceptual framework for iterative exploration in data-driven scientific discovery.}
    \textbf{(a-e)} Five-layer structure of the proposed framework, including user inputs, AI outputs (LLM-native figures and data-driven artifacts), and the underlying LLM engine and databases. 
    \textbf{(f)} Structure of an LLM-native figure at timestamp $t$, which supports bidirectional mapping between visual ($V$) and analytical ($C$ and $D$) representations, along with associated metadata $M$. 
    \textbf{(g)} Definition of a data-driven artifact. Starting from a focal figure, analyses can be iteratively refined or extended, supporting non-linear human exploration logic. Two example interaction modes are illustrated:
    (1) \textit{Manipulating} existing figures to refine or reorient the analysis focus, and
    (2) \textit{Extending} existing figures from user-interested data points to generate new, coordinated figures that enable progressive, drill-down exploration through filtering and brushing. 
    }
\end{figure}

\textbf{Bidirectional Mapping.}
To support reliable round-trips between visual and analytical representations of a figure, our framework maintains an explicit bidirectional mapping $R_t$ at time $t$ that links visual marks and graphical user interface (GUI) interactions in the visualization to the underlying analytical operations in $C_t$ and data rows in $D_t$ that produced them.
This mapping forms the basis for two complementary transformations: 

\begin{itemize}
    \item \textit{Analytical operations → Visualization.} 
    The LLM parses natural-language instructions into a sequence of analytical actions (Sec.~\ref{subsection:method-llm}-Action Space), including selecting relevant tables and columns, filtering rows, and assigning the chart type, visual encoding, chart interactions, and textual insights.
    These actions generate code $C_t$ and compute data $D_t$, resulting in a new or updated visualization $V_{t+1}$ with a set of visual marks (\eg, bars in the bar chart) to render the figure (Fig.~\ref{fig:Figure2}a, b: \textsquarenumber{1} → \textsquarenumber{2} → \textsquarenumber{3}).
    \wyf{In our example, after receiving the user instruction, 
    the LLM selects the \textit{temperature}, \textit{year}, \textit{month}, and \textit{state} fields, filters rows by \textit{state} and \textit{year} fields, and generates the code and data that produce Fig.~\ref{fig:Figure1}a.}
    \item \textit{Visualization → Analytical operations.} 
    When users interact directly with a visualization $V_t$, the system translates the visual interactions (\ie, the selected visual marks) into a new sequence of analytical operations. 
    Specifically, the system identifies the highlighted visual marks, retrieves the corresponding data rows, and combines the additional analysis requirements in the user's instruction to construct a new sequence of analytical actions and then generate a new visualization. 
    \wyf{In our example,  when a user brushes the time range June-August in Fig.~\ref{fig:Figure1}a and asks a follow-up question about the summer period (Fig.~\ref{fig:Figure1}b), the system filters the data to the selected period, constructs the new analytical operations, and generates a second figure} (Fig.~\ref{fig:Figure1}c, d, e; Fig.~\ref{fig:Figure2}a, b: \textsquarecharacter{a} → \textsquarecharacter{b} → \textsquarecharacter{c}).
\end{itemize}
This capability enables mixed-initiative workflows in which humans and LLMs alternate between specifying intentions in natural language and refining results through direct manipulation.

\textbf{Data-Driven Artifact and Provenance.}
To preserve the provenance of each insight and the user's iterative exploration logic, every analytical step, whether triggered by language or by visual interaction, passes through a single execution pipeline. 
The pipeline first interprets the user input ($I_t$), then generates and executes code ($C_t$) in a controlled environment, updates the data subset ($D_t$), and generates a new visualization ($V_{t+1}$).
The outputs of this process are recorded in a \textbf{data-driven artifact} (Fig.~\ref{fig:Figure1}i, Fig.~\ref{fig:Figure2}g), a reusable, composable unit of exploration that stores LLM-native figures, together with their analysis code, underlying data, and coordination rules. 
While an LLM-native figure captures a single analytical question, an artifact captures the trajectory of states across one or multiple figures and the coordination relationships that link them.
Each artifact records not only the resulting figures but also the sequence of exploratory logic. 
This sequence consists of two kinds of operations: (1) 
\textit{manipulations}, which refine or reorient the analysis either by updating an existing figure within the same artifact node through modifications to $C_t$, $V_t$, or $D_t$, or by forking an existing artifact into multiple parallel analysis artifacts that inherit and update the source artifact's $V_t$, $C_t$, and $D_t$;
and (2) \wyf{\textit{extensions}}, which generate new coordinated figures from user-selected data points, thereby extending the artifact and analysis logic.

Each artifact $A_t$ at timestamp $t$ consists of three components: \begin{itemize}
    \item \textbf{User input ($I_t$)}: the user's natural-language instruction or graphical interaction and its parsed structured representation (Fig.~\ref{fig:Figure2}a). 
    \item \textbf{Derived figures ($F_t$)}: one or more figures generated during the exploration of a research problem. Each figure includes the analytical execution result (code $C_t$ and data $D_t$), the resulting visualization $V_t$, and associated metadata $M_t$ (Fig.~\ref{fig:Figure2}f). 
    \item \textbf{Coordination linkage and schema}: the updated coordination graph and code that records how the figures within the same artifact are connected to each other through brushing or filtering interactions (Fig.~\ref{fig:Figure1}d, g). 
    When the user selects a new region of interest in a figure (Fig.~\ref{fig:Figure1}f), the data and visualizations for the linked figures automatically update to display the selected data subset. 
\end{itemize}

%

Each artifact $A$ is maintained as a version-controlled ledger ($A_{t_1}$, $A_{t_2}$, ... $A_{t_n}$), forming a directed acyclic graph of artifact versions in which each node represents a reproducible state and each edge an analytical transformation. 
This design provides two key benefits. 
First, it enforces \textit{reproducibility}. Because all analytical actions, code, and data used to generate a figure are explicitly recorded, each artifact preserves complete analytical provenance and therefore enables deterministic reconstruction of the original figure by re-executing its stored code against the underlying data.
Low-level differences in these reconstructions arise only when the original analysis involves non-deterministic procedures, such as stochastic analytical operations (Supplementary Note 4.3).
Second, it enables reasoning \textit{transparency}, allowing both humans and models to audit intermediate analytical steps (such as action-level reasoning and results) and explain results at different levels of detail. 
Artifacts thus serve as both memory and medium: a living record of discovery that can be edited, extended, revisited, and shared across similar analysis tasks. 

\textbf{Integration with Large Language Models.} 
Large language models operate as the core coordination layer through a plan–action–observation loop (Sec.~\ref{subsection:method-llm}).
The model receives user instructions and interactions, along with contextual information from the artifact and conversational history, to generate candidate analytical actions, reason over complex data, and validate them through code execution.
Execution outputs, including visualization and textual insights, generated data, and potential errors, are then fed back to the model to inform subsequent action selection. 
To constrain the model's behavior and ensure reliability, LLMs choose actions from a constrained \textit{action space} (Sec.~\ref{subsection:method-llm}) that defines permissible operations (\eg, filtering, transformation, modeling, and visualization). 
Operating within this space, the LLM translates natural-language input into valid analytical action sequences that are interpretable, executable, and reversible through provenance records.
This integration turns the LLM into an orchestrator of analytical operations rather than a black-box generator, aligning model reasoning with the explicit structure of the computational pipeline.
This framework also allows domain-specific customizations (\eg, tailored AI agents and tool designs~\cite{gao2025democratizing}) to be integrated into the system, adapting these capabilities to specialized domain contexts.

\subsection{Case Studies} 
\label{subsection:case-study} 
The climate example in Sec.~\ref{subsection:conceptual-framework} provides a simple illustration of the mechanics of our domain-independent conceptual framework. 
Here we demonstrate the utility of the proposed framework through~\systemname, an instantiation tailored to the science of science (SciSci) domain. 
SciSci provides an ideal testbed because it combines massive, heterogeneous datasets spanning grants, publications, and patents;
complex research problems, such as tracing knowledge evolution; and high-stakes decision-making scenarios, such as strategic funding allocation.
We present four case studies with a fifth in Supplementary Note 2.5 in which \systemname is used to explore SciSci challenges.

\subsubsection{Case Study 1: Innovation Landscape at Universities}
In this case, we use \systemname and its integrated innovation-related dataset from a leading U.S. research university (Supplementary Note 2.2). 
Understanding the research innovation landscape has become a central research issue in SciSci, where researchers aim to map the dual frontiers of science and technology~\cite{ahmadpoor2017dual, liang2022systematic, yin2022public}, and a concern for university leaders and science policymakers seeking to accelerate technology transfer and amplify research impact~\cite{wang2023innovationinsights, wang2025funding}. 
Yet uncovering untapped innovation potential remains challenging: relevant signals are distributed across heterogeneous data sources, and exploratory questions evolve rapidly as new insights emerge. 

Imagine a researcher seeking to explore the innovation landscape at the level of individual inventors. 
The researcher might first ask \systemname:
\textit{``Show me the distribution of inventors based on their number of invention disclosures (as y axis) and number of papers cited by patents (as x axis).''} 
The system generates a sequence of atomic actions (\eg, \actionname{select_table} and \actionname{add_chart_type}), executes the associated code, and returns an interactive scatterplot showing the inventor distribution at this university (Fig.~\ref{fig:Figure3}a), with the y-axis encoding invention disclosures (a proxy for present-day innovation activity) and the x-axis encoding papers cited by patents (a proxy for technological potential).

The figure shows that most inventors cluster tightly in the lower-left region due to the heavy-tailed nature of both measures. 
This compression obscures meaningful structure and hinders the identification of high-potential individuals. 
The researcher might then select the figure and request:
\textit{``Update the chart to use log scales on both axes. Customize the calculation to apply log(value + 1) so zero values are handled.''}
Here, the system shifts from generation to manipulation (Sec.~\ref{subsection:conceptual-framework}-Data-Driven Artifact and Provenance). 
The system takes the existing visualization and underlying data as inputs, infers the user's intent, triggers a new chain of actions to transform the data (\eg, \actionname{derive_column}) and update the visual encoding (\eg, \actionname{update_encoding}), and regenerates the scatterplot with log transformations applied to both axes, providing a clearer, more discriminative view of the inventor distribution (Fig.~\ref{fig:Figure3}b, c).
The overall distribution indicates a positive relationship between the two metrics, suggesting that patent-cited publications may serve as a useful indicator of a researcher's innovation activity.
Hover interactions expose precise values for each inventor, enabling accurate inspection of potential inventors and outliers (Fig.~\ref{fig:Figure3}b).

\begin{figure}[!t]
    \centering
    \includegraphics[width=1\linewidth]{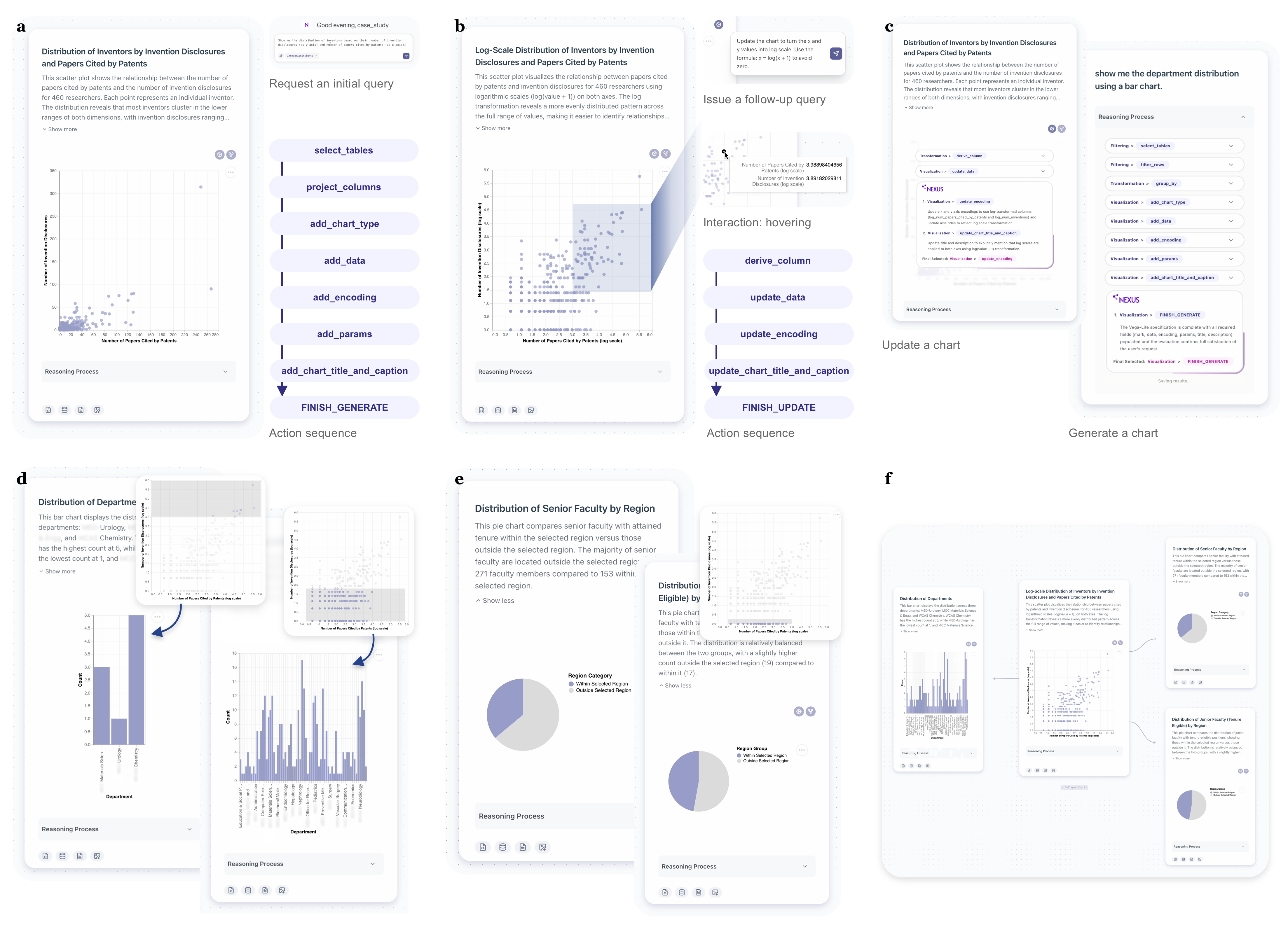}  
    \vspace{1pt}
    \caption{
    \bodyfigurelabel{fig:Figure3}
    \textbf{Case study: University innovation landscape.}
    \textbf{(a)} The figure generated for the first user query. 
    \textbf{(b)} The updated figure after adjusting the x- and y-axes to log scales. 
    \textbf{(c)} Screenshots of the streaming output when generating and updating a figure.
    \textbf{(d-e)} Comparison of different groups of researchers after filtering and coordinating across multiple figures. 
    \textbf{(f)} Screenshot of the data-driven artifact. 
    }
\end{figure}

\wyf{
The overall distribution reveals a small group of individuals with particularly high numbers of invention disclosures (Fig.~\ref{fig:Figure3}d, left scatter plot). 
To examine this group, the researcher might brush the corresponding region through direct manipulation on the scatterplot and pose a follow-up query to drill down: 
\textit{``show me the department distribution using a bar chart.''}
The system uses both the natural-language request and the GUI-based selection as joint inputs, triggering a new sequence of atomic actions. 
It then generates a bar chart summarizing the departmental composition of the selected group (Fig.~\ref{fig:Figure3}d, left bar chart), enabling quick assessment of departmental patterns among these highly active inventors.
By brushing different regions on the scatterplot and comparing the department composition of this group against those with few or no invention disclosures (Fig.~\ref{fig:Figure3}d, right scatterplot), the researcher would observe that faculty with frequent invention disclosures are disproportionately concentrated in fields such as chemistry and materials science, reflecting the uneven distribution of patenting activity across research areas within the university (Fig.~\ref{fig:Figure3}d, two bar charts).
}

\wyf{
The researcher might also notice an interesting group in the bottom region of the scatterplot (Fig.~\ref{fig:Figure3}e, scatterplot): these faculty have no disclosed inventions, yet their publications are cited by patents. 
This subgroup represents researchers with significant untapped innovation potential: their work has demonstrable technological relevance yet has not been translated into their own patent disclosures. 
To examine the relationship between this untapped potential and their tenure status~\cite{tripodi2025tenure}, the researcher could brush the corresponding points and ask analogous queries for senior and junior faculty, 
\eg, \textit{``show the distribution of all senior faculty (tenure `Attained'), comparing those within this selected region of researchers against those who are not, using a pie chart.''} 
Interestingly, nearly half of junior faculty fall within the selected region, compared to only about one-third of senior faculty (Fig.~\ref{fig:Figure3}e, two pie charts). 
This asymmetry points to a novel discovery.
Junior faculty, whose career advancement depends primarily on publications and grants, have strong incentives to prioritize academic output over other impacts such as direct engagement with patenting, yet their research is nonetheless absorbed by the technology sector, as evidenced by patent citations. 
Senior faculty, by contrast, are more likely to have established the industry collaborations and institutional networks that facilitate direct participation in invention disclosure and technology transfer.
The observed divergence suggests that a substantial reservoir of commercially relevant knowledge produced by junior faculty remains unrealized by the formal technology transfer pipeline, highlighting a structural gap between academic incentive systems and innovation outcomes that merits systematic investigation.
}

To ensure analytical rigor, the researcher could download the underlying data, generated code, and visualization outputs from each figure to verify the procedure and results. 
The entire chat session, including the data-driven artifact (Fig.~\ref{fig:Figure3}f), is also saved for future analysis, enabling replication and facilitating cross-institutional comparisons as additional university datasets become available. 

This case study illustrates how LLM-native figures support descriptive analysis and provenance-rich exploration. 
Unlike conventional tools that require manual reconstruction or produce static outputs, our approach records each analytical step as a versioned, navigable artifact, enabling mixed-initiative analysis that existing systems do not provide.

\subsubsection{\newlyadded{Case Study 2: Use of Science in Public Policy}}
\label{subsubsection:case2-science-policy}
\newlyadded{
The second case study explores the relationship between science and politics by tracing the scientific knowledge that informs policymaking.
Scientific evidence is widely regarded as a cornerstone of public policy, shaping legislative responses to complex societal challenges such as climate change and public health~\cite{furnas2025partisan, furnas2025partisanfunding}.
Prior research has shown that the uptake, framing, and interpretation of scientific evidence can be systematically shaped by the political context, particularly under conditions of sustained partisan polarization~\cite{mccarty2016polarized}.
This raises a fundamental question at the intersection of science and governance that can be examined with \systemname: do policymakers from different political parties draw on systematically different bodies of scientific knowledge?
}

\newlyadded{
To explore this question, a SciSci researcher could use \systemname and its integrated science policy dataset (Supplementary Note 2.3) to determine the kinds of science that are used in policy. 
As a tractable yet substantively rich starting point, the researcher might focus on documents from the U.S. House Committee on Energy and Commerce. 
The committee offers an analytically useful case because its science-intensive portfolio spans health, environmental regulation, and energy, generating citations to a heterogeneous range of scientific topics~\cite{furnas2025partisan}. 
The researcher could begin by asking: 
\textit{``Show a scatterplot of all the papers cited by House Energy \& Commerce Committee policies (committee short name is ``hsif00"), where each point is one unique paper title and doi and positioned by its title text embedding.''} 
\systemname first filters the policy corpus by source committee and extracts the cited papers along with their titles.
Next, the system applies two complex analysis actions (\actionname{text_embedding} followed by \actionname{dimension_reduction}) and generates a scatterplot in which each point represents a cited paper positioned by semantic similarity.
The figure shows clear topic clusters based on the visual distribution of the papers (Fig.~\ref{fig:Figure4}a). 
}

\newlyadded{
To examine the differences between paper topics cited by Democrats and Republicans, the researcher could issue another query: 
\textit{``Assign a dark blue color to a paper point if at least one House Energy \& Commerce Committee policy (committee short name is ``hsif00") citing that paper is under Democratic party control; otherwise, color it light grey. Reuse the existing scatterplot layout, and keep each paper's coordinates and the axis limits unchanged.''}
The system then takes the generated figure and the underlying data as input to query the original database, adding each paper-policy citation pair with a new variable from the dataset indicating party control. 
It updates the data in the figure as well as the visual encoding, and it returns an updated scatterplot with papers that are cited by policies written under Democratic control highlighted in blue (Fig.~\ref{fig:Figure4}b).}

\newlyadded{
Having identified visually distinct clusters, the researcher might also seek to understand the scientific content represented by each region. The researcher could then circle a cluster on the scatterplot and issue a follow-up request to summarize the paper topics: 
\textit{``Looking at the selected region and for papers cited by policies under Democratic party control, generate a topic label based on paper titles.''} 
The system queries the database for the papers in the circled region, retrieves their titles, and calls the LLM to determine the region's topic. 
The researcher might also ask the system to list five of these papers to confirm the accuracy of the topic label: \textit{``Looking at the selected region and for papers cited by policies under Democratic party control, provide a raw data table with each paper's title and doi. Return the first five paper records.''} 
It then returns a sample of the papers' metadata so that the researcher can cross-validate the inferred label.
The researcher might then brush several other visual clusters and record the representative topics: \textit{e-cigarette health effects, marketing, and regulation} (Fig.~\ref{fig:Figure4}c) and \textit{alcohol-impaired driving prevention and traffic safety} (Fig.~\ref{fig:Figure4}d). 
}

\newlyadded{
To examine papers cited in policy documents produced while the committee was under Republican control, the researcher might select the existing artifact for the Democratic-control analysis and request:
\textit{``Repeat the same analysis and artifact for the other party---this time using policies and papers associated with Republican control.''}
The analysis is nearly identical, with minor adjustments to the data query and visualization representation. 
The system forks the artifact, including its underlying code, data, and figure, and then determines the analysis and visualization information to be updated: it queries papers cited by policies produced while the committee was under Republican majority control, updates the visualization with the new set of papers, and changes the color to a shade of red associated with the Republican party, drawing on the LLM's general world knowledge. It also updates the figure title and description. 
The forked artifact for the Republican case then appears in the session. 
Following the same procedure, other representative topics would be identified, such as \textit{high school sports injuries and safety} (Fig.~\ref{fig:Figure4}e). 
}

\newlyadded{
This exploration suggests that Democrats and Republicans draw on distinct bodies of scientific research in policymaking, with direct implications for evidence-based governance. 
The finding was enabled by \systemname's integration of advanced data analysis, direct graphical capture of data points that cannot be precisely expressed in natural language, and rapid forking of complete analytical workflows---capabilities that remain difficult to achieve efficiently with conventional approaches.
}

\begin{figure}[!t]
    \centering
    \includegraphics[width=1\linewidth]{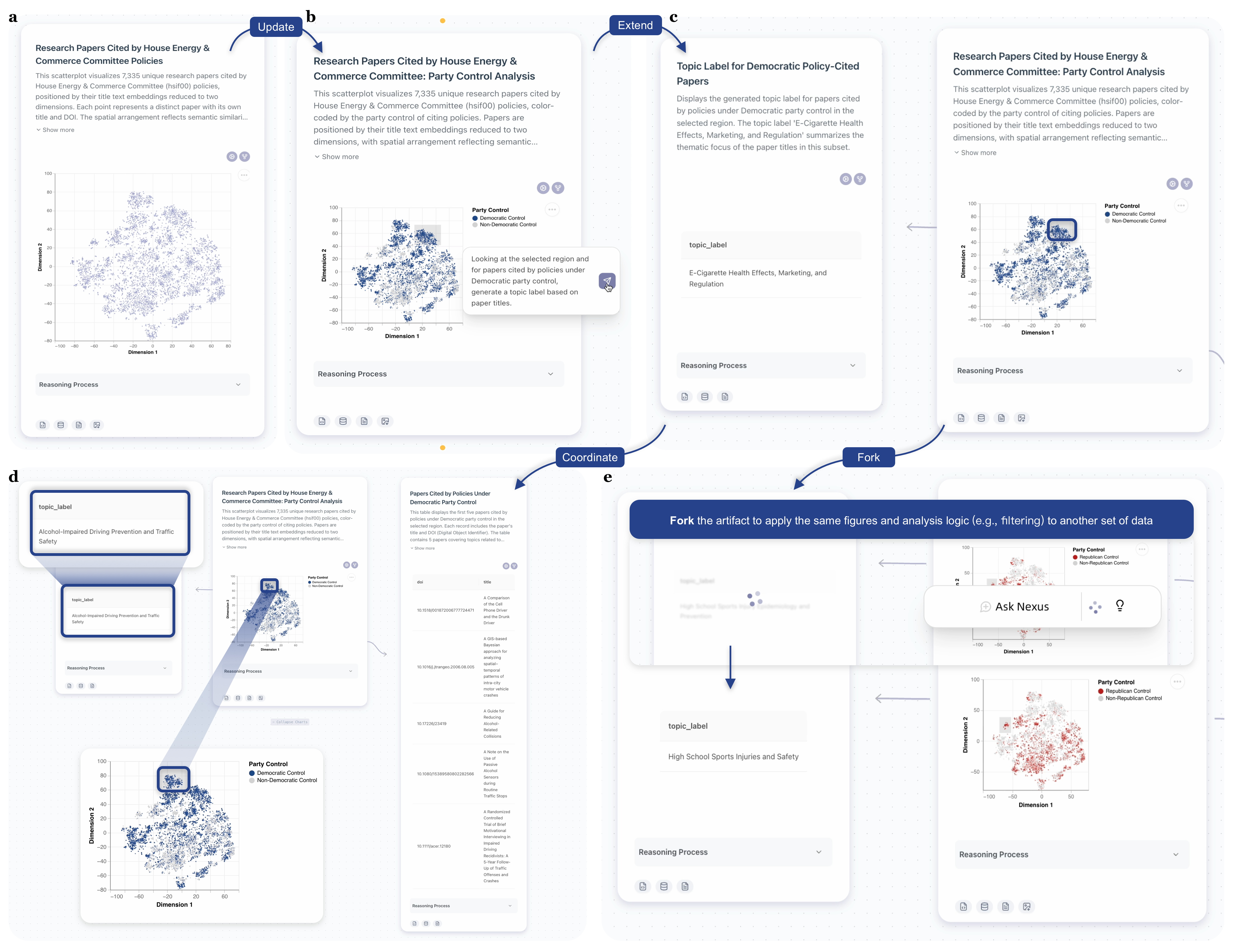}
    \vspace{1pt}
    \caption{
        \bodyfigurelabel{fig:Figure4}
        \newlyadded{\textbf{Case study: Use of science in public policy.} 
            \textbf{(a)} The scatterplot generated for the initial query, showing all papers cited by policy documents.
            \textbf{(b)} An in-place manipulation of the same figure highlighting papers cited by policy documents produced when the committee was under Democratic control.
            \textbf{(c)} An extension of the artifact: the user brushes a region of interest, and the system generates a coordinated figure summarizing the topic label of the selected paper cluster.
            \textbf{(d)} Coordination in action: changing the graphical selection region propagates to the linked topic summary, while the underlying data table remains accessible for cross-validating the generated labels.
            \textbf{(e)} The artifact forked from Democratic- to Republican-associated papers, re-executing the same analytical chain on a new data scope and revealing markedly different paper topic distributions across the two parties.
        }
    }
\end{figure}

\subsubsection{\newlyadded{Case Study 3: The Impact of Science Funding}}
\label{subsubsection:case3-funding-impact}
\newlyadded{
In the third case, we use \systemname and an integrated science funding dataset (Supplementary Note 2.4) to examine the funding portfolio of the National Institute on Aging (NIA), a major institute of the U.S. National Institutes of Health (NIH), tracing an analytical chain from funding allocation to representation to impact. 
Understanding the impact of science funding is critical to ensuring that research investments align with societal needs~\cite{bush2021science, yin2022public, wang2025funding}, a question of immense interest to funding agencies, national policymakers, and university leaders who allocate scarce resources across a complex research landscape. 
While emerging datasets now link grants to downstream outcomes well beyond publications, including patents, clinical trials, policy documents, and news coverage~\cite{yin2022public, wang2025funding, furnas2025partisan}, assessing this broad impact remains challenging: the evidence is scattered across heterogeneous databases, and portfolio-level questions about where funding goes naturally cascade into questions about who receives it and what impact it generates.
}

\newlyadded{
Here, a SciSci researcher could begin by asking \systemname for a portfolio overview: 
\textit{``For grants awarded by the National Institute on Aging, show the total funding amount by Level 0 research field, sorted from highest to lowest.''} 
The system searches for the funder, filters grants to those awarded by the NIA, joins each grant to its field classification, and aggregates funding by field, returning a ranked bar chart that spans multiple top-level research fields (Fig.~\ref{fig:Figure5}a, c). 
The portfolio is dominated by the four top-funded grant fields: \textit{biological sciences}, \textit{health sciences}, \textit{biomedical and clinical sciences}, and \textit{psychology}.
}

\newlyadded{Noticing this concentration, the researcher might turn from the funding to the recipients, asking who leads the research these grants support. 
To investigate, the researcher could brush the bars for the three top-funded fields using direct manipulation (Fig.~\ref{fig:Figure5}a) and pose a follow-up query: 
\textit{``Compare the gender shares of principal investigators within these selected fields using a grouped bar chart, and include an additional ``Overall" group showing the gender share across all fields.''} 
Taking the brushed region and the natural-language request as joint inputs, the system maps the selected bars back to their field conditions and inherits the NIA scope from the parent figure, so that every downstream comparison remains within the same portfolio. It then computes the share of distinct principal investigators (PIs) by gender within each field and constructs the requested ``Overall" baseline across all NIA grants.
The resulting grouped bar chart (Fig.~\ref{fig:Figure5}b) reveals a consistent imbalance: male PIs outnumber female PIs in every top-funded field, with the disparity in \textit{biological sciences} and \textit{biomedical and clinical sciences} exceeding the overall gap.
}

\newlyadded{
Given this finding regarding representation, the researcher might shift their focus to the question of research impact, asking whether the impact of grants led by female PIs differs from that of grants led by male PIs. 
To investigate this question, the researcher might select the single bar representing \textit{biomedical and clinical sciences} and issue a computationally precise request:
\textit{``For this selected field, use a radar chart to compare male and female PIs' average grant-level impact. Use the precomputed impact metrics (number of publications, patents, clinical trials, and citations from policy documents, news, patents, and clinical trials). Average each metric over PI-grant records within each gender, then normalize each dimension by the maximum across the two genders.''} 
The system translates this specification into the corresponding aggregation pipeline: 
it averages the seven precomputed impact measures over PI–grant records within each gender, applies the normalization, and renders an overlaid radar chart (Fig.~\ref{fig:Figure5}d). 
While female PIs are underrepresented in this heavily funded field, the chart shows that their grants perform at least as well on many of the impact measures examined here: average publication output is nearly identical, and grants led by female PIs record higher average citations from policy documents, news media, and clinical trials---a pattern that prior analyses had not surfaced.
}

\newlyadded{
Although this finding is descriptive, the asymmetry between representation and realized impact bears directly on evidence-based funding allocation and merits systematic investigation---for example, controlling for career stage, funding scale, and team composition.
Notably, conventional gender-gap analyses, which center on publications, citations, and funding amounts~\cite{lariviere2013bibliometrics,huang2020historical,oliveira2019comparison}, would not surface this pattern. 
It becomes visible only when grants are connected to their downstream societal uses through newly available data linkages~\cite{yin2022public, wang2025funding}. \systemname helps the researcher to extract a concrete and intuitive visual finding from these complex raw tables in a short sequence of intuitive and efficient interactions.
}

\begin{figure}[!t]
    \centering
    \includegraphics[width=1\linewidth]{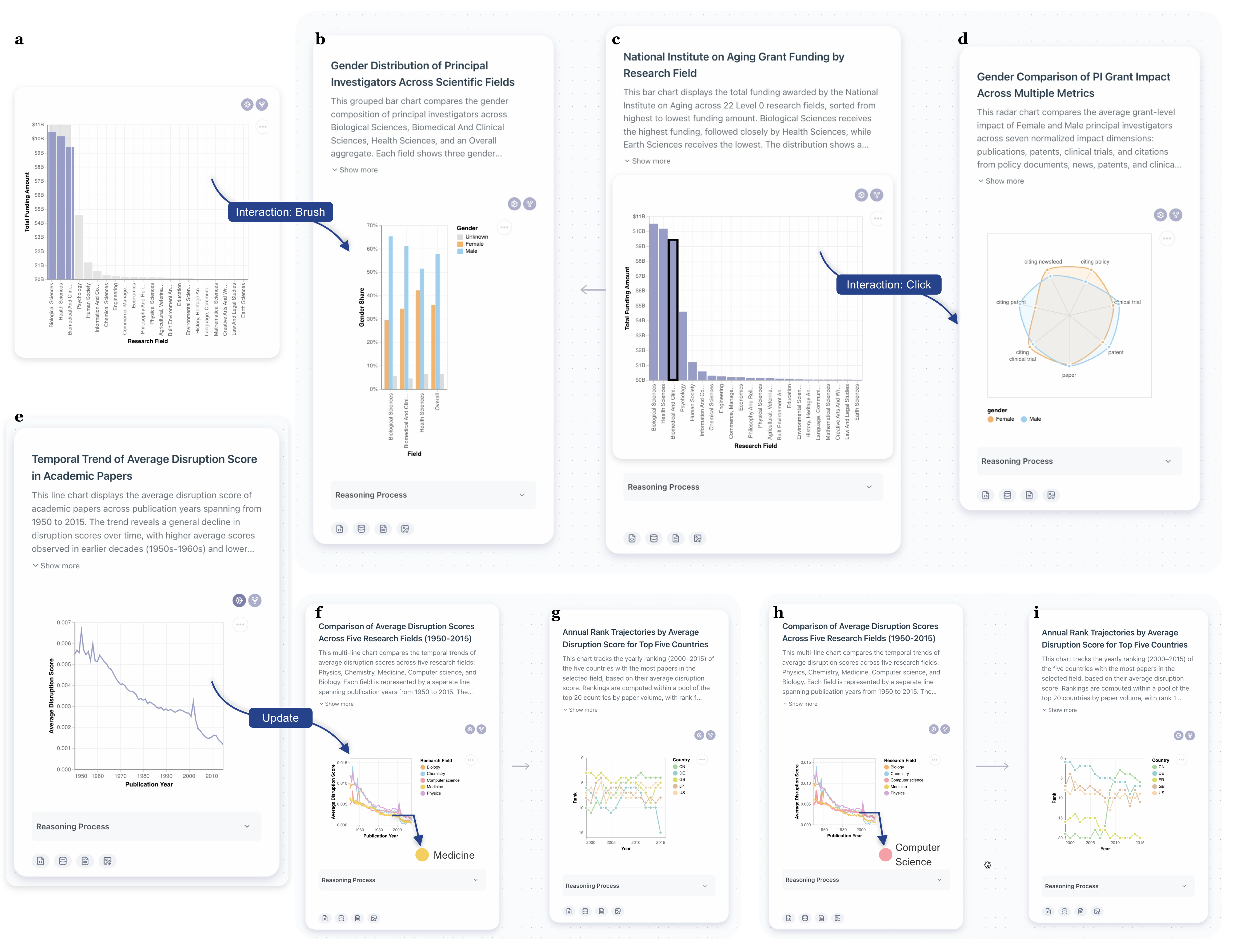}  
    \vspace{1pt}
    \caption{
        \bodyfigurelabel{fig:Figure5}
        \newlyadded{
            \textbf{Case studies:}
            \textbf{(a-d) Impact of science funding.} 
            Starting from the distribution of grant funding across research fields \textbf{(a, c)}, the researcher could compare the gender composition of principal investigators (PIs) within the brushed fields \textbf{(b)}, then zoom into a single field to contrast the multidimensional impacts of male- and female-led grants in a radar chart \textbf{(d)}.
            Color encodings are designed to be semantically aligned across all figures within the same data-driven artifact.
            \textbf{(e-i) Disruptive science across fields and nations.}
            Beginning from a replication of a known macro-level pattern~\cite{park2023papers} \textbf{(e, f)}, new insights emerge as the analysis is drilled down to the relative ranking of disruption scores across countries \textbf{(g-i)}.
        }
    }
\end{figure}

\subsubsection{\newlyadded{Case Study 4: Disruptive Science Across Fields and Nations}}
\newlyadded{
The final case study explores whether science is losing its capacity to disrupt, which has become a central concern in the SciSci community, fueling debates over whether scientific progress is slowing despite the exponential growth in the scientific literature~\cite{jones2009burden, park2023papers, wu2019large}. 
To examine this question, a SciSci researcher could use \systemname and the integrated SciSciNet dataset~\cite{lin2023sciscinet} (Supplementary Note 2.4) in a reproduce-then-extend exploration, first replicating established macro-level findings and then carrying the same analytical chain into largely unexamined dimensions.
}  

\newlyadded{The SciSci researcher might begin this exploration with an aggregate view: \textit{``Show the average disruption score of papers by publication year from 1950 to 2015.''} 
The system aggregates paper-level disruption scores by publication year and returns a line chart (Fig.~\ref{fig:Figure5}e), which shows that the average disruption declines persistently across the entire period. 
The researcher could then drill down into the analysis: 
\textit{``Split this line into multiple lines to compare the top five research fields (Level 0) with the most papers.''} 
The system shifts from figure generation to manipulation: it identifies the five fields with the most papers, augments the existing aggregation with a field grouping, and updates the visual encoding in the figure, transforming the single trend into five field-level trajectories (Fig.~\ref{fig:Figure5}f). 
All five fields decline, though at different rates: physics and chemistry, initially the most disruptive, fall most sharply, while computer science declines more gradually, rising in relative rank from last to near the top.
}

\newlyadded{This field-level heterogeneity and the aggregate decline replicate patterns documented in prior work~\cite{park2023papers}. Whether these trends extend uniformly across nations, however, remains largely unexamined in this literature. 
The researcher might therefore move from replication to new analysis by selecting the line for \textit{medicine} and posing a computationally precise follow-up request: 
\textit{``For the selected field, from 2000 to 2015, compare countries by average disruption score, counting each paper once per country (deduplicate multiple authors from the same country). Then among the 20 countries with the most papers in the selected field over this period, rank the countries within each year by average disruption score (rank 1 means the highest; ranks range from 1 to 20 within the 20-country pool). Show the annual rank trajectories of only the five countries with the most papers, as a multi-line chart.''}
The system executes the corresponding multi-step analysis: it restricts the sample to the twenty largest producers, computes within-year ranks, plots trajectories for the five highest-volume countries, and returns a rank-trajectory chart (Fig.~\ref{fig:Figure5}g). 
The resulting chart shows that national standings in this field evolve only gradually. The five largest paper producers remain in the upper half of the twenty-country pool throughout the period, with the United States and China trending slowly upward and the United Kingdom and Japan trending slowly downward over this period.
Selecting another field and repeating the request re-executes the same pipeline with only the field condition changed. For \textit{computer science}, however, the result is strikingly different (Fig.~\ref{fig:Figure5}h, i): 
China rises from near the bottom of the twenty-country pool in the early to mid-2000s to the top ranks between 2008 and 2011, remaining there through 2015 while several incumbent leaders gradually decline.
}

\newlyadded{
These trajectories suggest that beneath the universal decline in average disruptiveness lies an actively shifting national landscape, and that its dynamics are field-specific.
As concerns about the declining disruptiveness of science mount, the ability to track not only whether disruption is falling but also where and for whom it is rising or eroding can directly inform science policy debates about research portfolios and national competitiveness.
This case study thus illustrates a reproduce-then-extend mode of exploration enabled by \systemname: a published macro-level pattern is first re-established as a verifiable starting point and then progressively decomposed, first by field and then by nation, through figure extension and user interactions. 
}

\subsection{\newlyadded{User Study}}
\label{subsection:user-study} 
\newlyadded{To assess the potential benefits of LLM-native figures and data-driven artifacts for researchers, we conducted a within-subject user study in which SciSci researchers interacted with \systemname, our instantiation of the design. 
The goal of the study is to examine how researchers use the system in practice to conduct exploratory analyses. 
Participants performed two exploratory data analysis tasks adapted from the case studies above with comparable difficulty, using \systemname for one and a baseline representing their conventional workflows for the other. 
We measured task completion time and performance, perceived usability, perceived workload, and perceived value of the system's key components. 
We also collected researchers' qualitative feedback on how working with \systemname affected their use of figures and their scientific discovery process. 
}

\subsubsection{\newlyadded{Participants and Study Procedure}}
\label{subsection:participants-and-study-procedure} 
\newlyadded{\textbf{Participants.} 
We recruited 12 SciSci researchers (P1--P12) through professional networks, including 5 women and 7 men, aged 24--45 years ($M = 31.3$). 
Participants spanned multiple career stages: 2 predoctoral researchers, 3 PhD students in their third year or later, 5 postdoctoral researchers, and 2 research assistant professors. 
Their experience with quantitative data analysis ranged from 3 to 10 years. 
All participants routinely analyze large-scale SciSci data and reported prior experience using LLM-based coding assistants (\eg, Claude Code~\cite{claudecode}) for code generation and for creating and editing visualizations. 
None of the participants was involved in the development of \systemname or had used it before the study. 
}

\newlyadded{\textbf{Baseline.}
The baseline system had the same computational setup as \systemname, including the server, coding environment, and database access. 
Researchers could use their preferred conventional tools (e.g., VS Code and Jupyter notebooks) and programming languages, and they had access to LLM assistance through Claude Code~\cite{claudecode}, allowing them to follow their typical data analysis workflows.
We used Claude Sonnet 4.5 in the baseline, the same model that powers \systemname, so that differences between the two conditions could be attributed to the system design rather than to the underlying models' capabilities.
}

\newlyadded{\textbf{Study Procedure.} 
Each study session was conducted remotely via Zoom with screen sharing and lasted approximately two hours. 
Participants first completed a consent form and a demographics questionnaire. 
We then introduced the project background and provided a 15-minute \systemname tutorial using the climate change example from Sec.~\ref{subsection:conceptual-framework}.
Researchers could ask questions throughout the tutorial to clarify their understanding. 
We also helped them set up the baseline environment and gave them time to configure it to match their preferred data analysis workflows.
Next, we asked researchers to reproduce two exploratory data analysis tasks adapted from the case studies (Sec.~\ref{subsubsection:case2-science-policy} and Sec.~\ref{subsubsection:case3-funding-impact}). 
For each task, we provided natural-language data analysis requirements, reference chart types~\cite{ren2017chartaccent, ren2018reflecting}, and the data schema and descriptions of the underlying database. 
Each task includes a list of exploration questions, collectively covering core system features, such as figure generation, extension, and coordination analysis.
Each participant completed both tasks, using \systemname for one and the baseline system for the other. 
To fully counterbalance task-condition assignments and condition order, we used all four possible combinations (e.g., Task 1 with \systemname followed by Task 2 with the baseline, or the reverse) and assigned three participants to each.
In each task, participants worked through the steps at their own pace and were encouraged to think aloud. 
After each task, participants filled out a questionnaire to evaluate the tool's usability and a questionnaire to assess their perceived workload. 
After completing both tasks, participants answered another questionnaire on the practical value of the tool and user experience of \systemname's core features. 
We then conducted a semi-structured interview about their experiences, comparisons with their conventional workflow, and the tool's potential practical impact (Supplementary Note 3). 
This study was approved by the Institutional Review Board of Northwestern University. 
}

\subsubsection{\newlyadded{Results}}

\newlyadded{\textbf{Task completion time and performance.} 
As shown in Fig.~\ref{fig:Figure7}a, participants completed the tasks substantially faster with \systemname ($M = 16.4$ min, $SD = 2.6$) than with the baseline ($M = 31.4$ min, $SD = 14.2$; exact Wilcoxon signed-rank test, $W = 3$, $P = 0.002$, $r = 0.92$). 
%
Each of the four steps per task was scored by a human expert and LLM judges as correct ($1$), partially correct ($0.5$), or incorrect ($0$), for a maximum of $4$ per participant and task. 
With \systemname, all 12 participants reached the full score on their task ($M = 4.0$). 
Two participants conducted a second attempt at one step: one submitted an incomplete follow-up prompt, while the other selected three bars by clicking rather than brushing, initially producing an incorrect chart. 
The latter exposes a design limitation: when a chart supports multiple ways to make the same selection, the system should provide brief, chart-specific interaction guidance.
%
Baseline performance was also high on the Funding Impact task ($M = 3.8$, $SD = 0.6$) but substantially lower on the Science Policy task ($M = 2.6$, $SD = 0.5$). 
Baseline errors concentrated in the steps that depend on a visual selection: in the Science Policy task, the three region-based steps were not accurate, because participants had to translate the circled region into numeric t-SNE ranges, either by manually typing axis readings or by letting Claude Code read a screenshot.
Only one participant was scored $1$ on these steps, who built an interactive chart for box-select.
The remaining errors were isolated slips typical of ad hoc data analysis: incomplete de-duplication of records, an overly permissive filter that pulled in unintended records, input fields other than those specified, and aggregation at the wrong unit of analysis.
}

\newlyadded{\textbf{Perceived usability}.
We assessed the tool's perceived usability using the System Usability Scale (SUS)~\cite{brooke1996sus}. 
Researchers rated \systemname as substantially more usable than the baseline ($M = 89.4$, $SD = 10.7$ for \systemname; $M = 64.8$, $SD = 20.1$ for the baseline).
Eleven of the twelve participants gave \systemname a higher usability score than the baseline. Eleven also rated \systemname above the conventional benchmark of 68~\cite{sauro2011practical}, compared with five who rated the baseline above this threshold.
}

\newlyadded{\textbf{Perceived workload.}
Study results indicate that \systemname reduced participants' perceived workload.
We used the NASA Task Load Index (NASA-TLX)~\cite{hart1988development} to estimate the users' perceived cognitive load. 
Participants reported significantly lower workload in the \systemname condition ($M = 1.9$, $SD = 0.8$) than in the baseline condition  ($M = 3.2$, $SD = 1.3$), based on the average score of the six NASA-TLX items (Wilcoxon signed-rank, $W = 7$, $P = 0.009$, $r = 0.82$).
At the item level (Fig.~\ref{fig:Figure7}b), \systemname received significantly lower ratings than the baseline for mental demand ($P = 0.012$), physical demand ($P = 0.004$), and frustration ($P = 0.008$).
Differences in temporal demand, performance, and effort were not statistically significant. 
These results suggest that \systemname substantially reduces the perceived cognitive and physical burden of conducting the analysis tasks and makes the overall process less frustrating.
}

\newlyadded{\textbf{Perceived value of \systemname's design components.} 
We also analyzed participants' ratings of the usefulness of nine \systemname design components (Fig.~\ref{fig:Figure7}c) with a 5-point scale (1 = not at all useful; 5 = extremely useful). 
All nine design components received positive ratings, with a mean above $4$ for every component. 
Bidirectional mapping and data-driven artifacts received the highest usefulness ratings, while the two data-management components also received high ratings. 
Within the multi-agent engine, the workflow itself received high ratings overall ($M = 4.5$, $SD = 0.9$). However, one researcher rated it a 2, explaining that the multi-agent workflow was largely invisible to the user and the user could not directly perceive how the module operated. 
Notably, this participant nevertheless found the visible action-level reasoning useful.
Action-level reasoning received the lowest usefulness rating, although it remained positive ($M = 4.2$, $SD = 0.8$). 
Researchers generally found the streamed sequence of actions more structured and interpretable than a raw chain of thought, but preferences for the level of detail varied. 
Some wanted concise summaries of the analytical rationale behind key steps, while others preferred minimal information resembling a lightweight progress indicator rather than an explicit reasoning trace.
}

\begin{figure*}[t]
    \centering
    \includegraphics[width=1\linewidth]{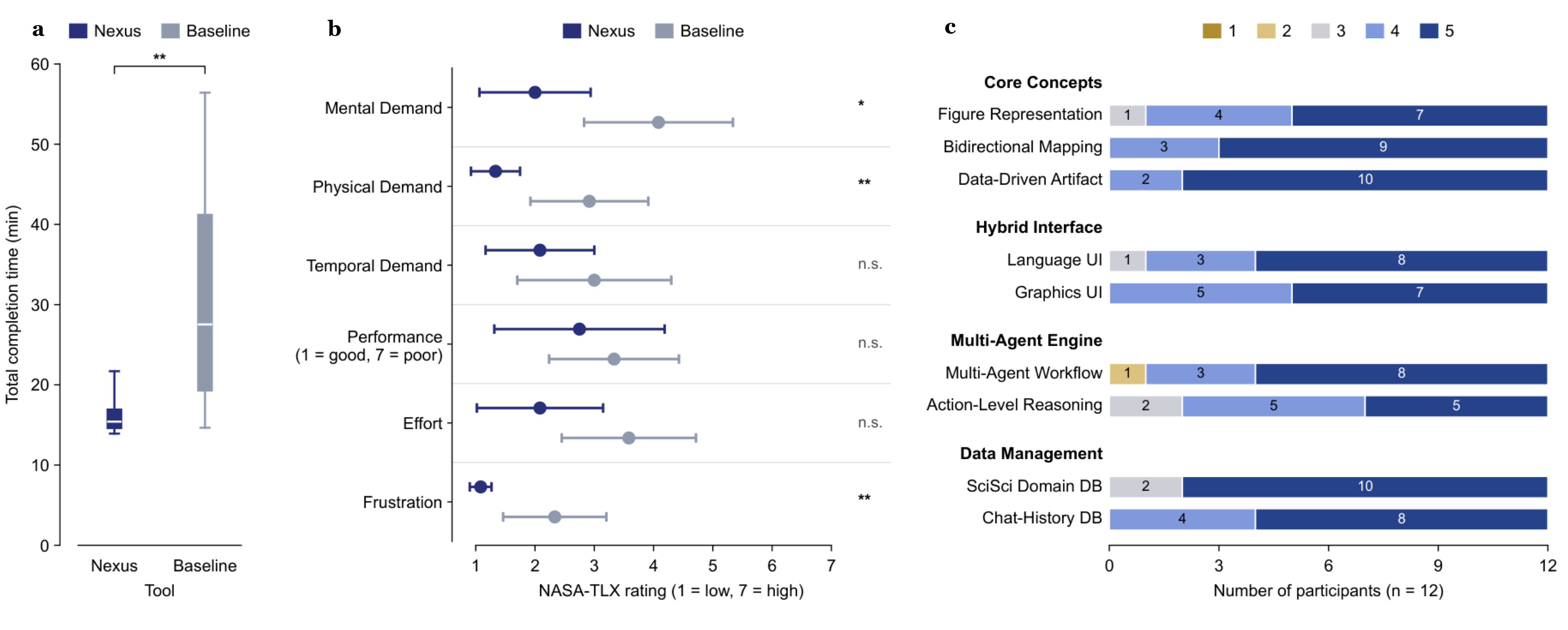}
    \vspace{1pt}
    \caption{
       \bodyfigurelabel{fig:Figure7}
       \newlyadded{\textbf{User study results.} \textbf{(a)} Task completion time with \systemname and the baseline. \textbf{(b)} Perceived workload by sub-scale. \textbf{(c)} Perceived value and usefulness of \systemname's core design components, grouped by framework module.}
    }
\end{figure*}

\newlyadded{\textbf{Qualitative findings.}
In addition to the quantitative results on efficiency and workload, semi-structured interviews and think-aloud observations across the twelve sessions revealed three broad ways in which \systemname shaped scientific discovery. 
}

\newlyadded{\textit{(1) Figures as interfaces for iterative discovery.}
Direct manipulation of interactive figures transformed analytical operations that were awkward, inefficient, or infeasible in the baseline into lightweight visual interactions. 
All six researchers who completed the science policy task with the baseline had to approximate regions of interest using coordinate ranges, annotated screenshots, or repeated prompting.
In \systemname, researchers could instead select visual marks directly and carry those selections into subsequent analyses, reducing both interaction overhead and ambiguity in specifying analytical intent. 
Participants particularly valued the ability to derive new coordinated analyses from an existing figure and selection, replacing analytical steps that otherwise required manually modifying and rerunning code.
}

\newlyadded{More importantly, LLM-native figures changed how researchers explored data. 
Direct manipulation of these interactive figures lowered the cost of inspecting regions, subgroups, and relationships, while also prompting analyses that participants said were absent from their usual workflows. 
Researchers zoomed in on local clusters, examined subpopulations and unanticipated gender differences, and proposed extensions incorporating new variables or subgroup comparisons. 
By making emerging questions easier to pursue, interactive figures allowed new directions of inquiry to arise directly from patterns encountered during exploration.
}

\newlyadded{\textit{(2) Data-driven artifacts as externalized reasoning trace.} 
Researchers repeatedly contrasted the data-driven artifact with the linear histories of computational notebooks and chats, in which branching analyses are compressed into a sequence, and the relationships among them must be maintained mentally. 
Several described the artifact as a ``mind map'' or ``mind-space'': by externalizing alternative branches, intermediate results, and their relationships, the artifact reduced the cognitive burden of tracking a project and made it easier to resume earlier directions or selectively extend particular branches. 
This externalized structure also supported communication across time and among collaborators; for example, participants saw value in using the artifact to reconstruct the reasoning behind previous decisions during advisor or team meetings, rather than rediscovering lost ideas between sessions.
Participants also valued the mechanisms that preserved this reasoning trace. For example, version rollback allowed researchers to revisit earlier states without contaminating the current conversational context. 
They suggested future extensions to support richer reasoning structures, including parallel branches beyond subgroup drill-downs, multiple root figures within an artifact, merged branches, and text-based intermediate analyses as first-class nodes. 
They also wanted the ability to delete nodes and to record each brushing or coordination state as a revisitable version.
}

\newlyadded{More broadly, preserving a non-linear exploration history opens the possibility of carrying scientific reasoning across sessions, collaborators, and ultimately into the scholarly record itself. 
Papers typically preserve the outcomes of discovery as a coherent retrospective narrative while leaving much of the intellectual trajectory behind them undocumented: which prior work first inspired an idea, how initially separate observations became connected, why some directions were abandoned, and which unexpected findings redirected the inquiry. 
Data-driven artifacts could make this otherwise ephemeral layer of science persistent, inspectable, and shareable. 
Externalized exploration traces could therefore provide not only a richer record of how science develops, but also a new substrate for grounding future AI systems in the process of scientific discovery rather than only in its published outcomes.
}

\newlyadded{\textit{(3) Provenance as basis for reproducibility and sharing.} 
Because every figure carried its data and code, verification became part of the analysis rather than a separate recovery step. 
Researchers identified three broader uses for these artifacts. 
First, for replication and open science, participants saw value in preserving downloadable code to support reproducibility and requested additional features such as automatically generated replication packages and exportable runnable notebooks for inclusion in supplementary materials. 
Second, for portability, they valued the ability to move an artifact across sessions and reopen it later without losing the computational context accumulated during a long conversation. 
They also proposed allowing users to copy an artifact into a new session while preserving the project's full exploration trajectory. 
Third, for science communication, participants envisioned publishing interactive artifacts alongside papers so that audiences could inspect subgroups, analytical paths, and supporting evidence, thereby making findings more accessible and increasing trust in the exploration process.
At the same time, participants emphasized that raw code alone was insufficient to establish trust, and they differed in the types of evidence they wanted to examine. While some requested analytical step summaries, others preferred access to the underlying data or independent cross-tool verification. 
These differences suggest that provenance mechanisms will need to adapt to researchers' heterogeneous verification strategies. 
}

\newlyadded{\textbf{Design suggestions.}}
\newlyadded {Participants identified several opportunities to improve \systemname and extend its capabilities. The most common request was support for analyzing researchers' own or locally stored datasets with \systemname, which several participants viewed as the primary remaining advantage offered by general-purpose coding agents over \systemname. 
Participants also asked for greater interaction flexibility, including natural-language control over visual styling, style transfer from reference figures, parallel analyses and multiple root charts, annotations attached to nodes, notebook or full-package export, and lightweight interaction guidance. 
They further requested more expressive graphical selection mechanisms, such as persistent brushes across queries, multiple concurrent selections, and a natural-language fallback when built-in interactions were insufficient. 
Finally, several participants requested greater visibility into multi-step reasoning and faster system responses.
}

\subsection{Computational Evaluation}
\label{subsection:computational-evaluation} 
To further assess the feasibility of LLM-native figures, we use \systemname to evaluate the fidelity of the bidirectional mapping mechanism (Sec.~\ref{subsection:conceptual-framework}), the core design for figures to function as reliable computational interfaces.
Specifically, we test whether: 
(1) user questions can be accurately transformed into visual data insights (\ie, Analytical operations → Visualization), and
(2) visual interactions, such as clicking or brushing on a figure, can be accurately mapped back to the intended subsets of underlying data and the appropriate analytical operations (\ie, Visualization → Analytical operations). 


\textbf{Test Design}.  
We adopt a structured, coverage-oriented evaluation strategy that treats mapping fidelity as a functional correctness problem over a well-defined interaction space. 
Using the datasets from our Case Study 1, we construct a test suite of paired queries. Each test case includes an initial user question that requires the generation of a figure, and a follow-up question that depends on a specific visual interaction with that figure to produce a new or updated visualization. 
This design mirrors realistic usage patterns in which users alternate between language-based queries and direct manipulation.

To ensure systematic coverage and reduce evaluator bias, we use an LLM to generate  \wyf{308} valid test cases, each with three sub-tasks (Supplementary Note 4.1). 
Test cases are diversified along three dimensions: 
(1) figure type (\eg, bar charts, line charts, pie charts, scatter plots, and tables),
(2) interaction type (\eg, single-mark selection, one-dimensional interval brushing, and two-dimensional brushing), 
and (3) analytical complexity, with tier 1 cases representing simple scenarios involving a single table and minimal transformation, and tier 2 cases reflecting realistic workflows that require joins across multiple tables, aggregations, or modeling operations.
%
We evaluate mapping fidelity along both directions of the language–visualization loop using three metrics across all test cases.
\textit{Execution Success Rate} is the proportion of test cases in which the system successfully completes the end-to-end execution and returns a usable visualization artifact (i.e., no execution failure and a chart is generated).
\textit{Conditional Accuracy} (conditional on success) is the proportion of cases in which the generated analytical process and results are semantically and logically correct with respect to the user's intended analysis.
\textit{End-to-End Accuracy} is the proportion of cases in which the figures are both successfully generated and analytically correct.

\begin{figure}[!htb]
    \centering
    \includegraphics[width=0.7\linewidth]{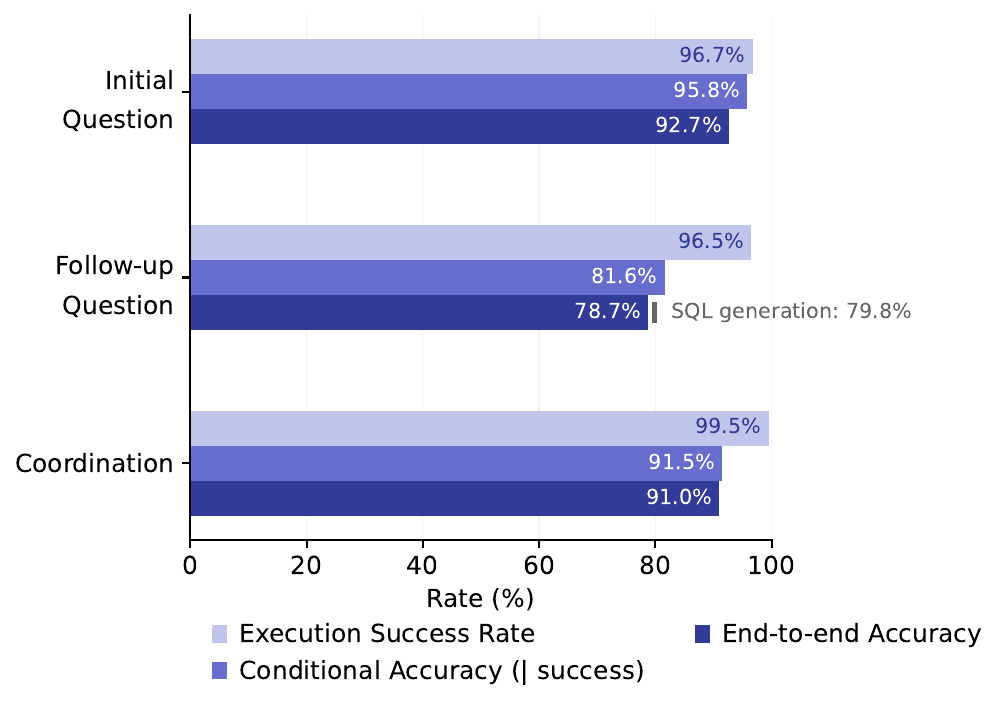}
    \vspace{1pt}
    \caption{
       \bodyfigurelabel{fig:Figure6}
       \textbf{Computational evaluation results.}
    }
\end{figure}

\textbf{Results.}
We first assess the mapping from analytical operations to visualization using the initial question in each test case. 
Given a user instruction, \systemname generates a sequence of actions to retrieve and analyze data and produce an interactive figure. 
We evaluate the correctness of the generated code (SQL, Python, and Vega-Lite), the retrieved data subset $D$, and the resulting figures through manual inspection by two authors and cross-validation with the generated LLM-based solution (Supplementary Note 4.2).
\systemname achieves an overall \textit{Execution Success Rate} of \wyf{$96.7\%$} and \textit{End-to-End Accuracy} of \wyf{$92.7\%$} (Fig.~\ref{fig:Figure6}-Initial Question). 
Most inaccurate cases stem from SQL generation errors, such as failures to perform fuzzy matching for user-specified concepts (\eg, querying “Computer Science” verbatim when the database stores ``computer science'') and misidentification of the analytical entity (\eg, returning counts of researchers instead of papers). 
%
%

Second, we assess the mapping from visualization back to analytical operations using two approaches: follow-up question-answering and inter-figure coordination. 
In the follow-up question-answering task, for each figure–interaction pair, \systemname maps selected visual marks to a data query via the relation $R_t$ and generates the corresponding analytical actions. 
We examine the results of the generated SQL query $C$ and the retrieved data subset $D$. 
When users select visual marks or regions on the figure, the system correctly infers the corresponding data filtering and transformation logic with \wyf{$79.8\%$} accuracy (Fig.~\ref{fig:Figure6}-Follow-up Question, SQL generation). 
We observe that occasional inaccuracies arise when translating GUI interactions (\eg, brushing or clicking) into SQL filtering logic.  
For instance, a continuous x-axis brush with a range of $[a, b]$ may be incorrectly mapped to a discrete ``IN'' predicate rather than a ``BETWEEN'' range condition. 
This problem highlights the need for more explicit guidance, such as structured prompts or interaction-aware constraints, to ensure that GUI-derived filtering semantics are faithfully translated into SQL predicates in future LLM system designs. 
For inter-figure coordination, when users select or brush an updated region of interest, the system updates the coordinated figure by modifying the underlying SQL query to retrieve the corresponding data. 
The resulting \textit{End-to-End Accuracy} is \wyf{$91.0\%$} (Fig.~\ref{fig:Figure6}–Coordination).

\section{Discussion}
\label{section:discussion}
Figures have long served as the primary medium through which scientists interpret and communicate data.
Yet, in most computational workflows, they remain detached from the analytic processes that generated them. 
The framework introduced in this paper demonstrates that this separation is not intrinsic. 
By embedding full provenance within visual artifacts, figures can become integral components of the computational system itself. 
This transformation shifts the figure's role from a static representation to a dynamic interface through which both humans and machines can reason about data.

Embedding provenance at the artifact level also redefines what it means for a visualization to be interpretable. 
For humans, interpretability derives from visual pattern recognition; for machines, it emerges from structured access to the underlying code and data.
By integrating these modalities, LLM-native figures enable compositional interpretability, in which a figure's meaning can be simultaneously read, executed, and verified. 

Despite its importance, provenance capture has been a persistent challenge in computational science. 
Conventional notebooks and code can track computational steps and data, but they often fail to connect these components to the human-facing outputs through which results are interpreted. 
The data-driven artifact addresses this gap by linking every figure to an executable lineage of analytical actions. 
This not only ensures reproducibility but also enables post hoc transparency: each step can be revisited, audited, or reinterpreted in context. 
In practice, this means that LLM-native figures can serve as a new unit of computational provenance. 
Rather than preserving entire analysis environments or code scripts, researchers can archive and share the figures themselves, each serving as a self-contained record of data, code, and visualization. 
Such artifacts can be queried, recombined, or regenerated as analytical building blocks, reducing the friction between exploration, publication, and verification.

Beyond improving provenance and reproducibility, the framework highlights a distinct form of complementarity between human and machine reasoning. 
Humans excel at recognizing salient visual patterns, while language models excel at executing systematic transformations and tracing dependencies.
LLM-native figures provide a shared representational substrate that allows each to operate within its comparative advantage. 
The human can identify an unexpected visual pattern; the model can immediately trace its provenance, recompute subsets, or test alternative solutions. 
This complementarity suggests a broader reframing of how computational systems might support scientific reasoning. 
Rather than viewing LLMs as autonomous analysts or assistants, the framework positions them as provenance-maintaining collaborators--agents whose primary function is to ensure that each analytical operation remains legible, reproducible, and extensible.

At a broader level, this perspective also points to a shift in how humans collaborate with generative AI. 
For generative AI, advancing beyond the dominant linear, end-to-end question–answering paradigm is essential for deeper alignment with human reasoning.
Rather than confining intelligence to opaque conversational outputs, future systems could externalize intermediate analytical reasoning as inspectable and revisable processes, shifting interaction from linear chat histories to artifact-based, non-linear exploration structures, where humans can revisit states, branch alternatives, and iteratively refine understanding in ways that more closely mirror how human knowledge is formed. 
Within such structures, generative AI can evolve from a reactive respondent into a collaborative reasoning partner that is able to trace exploration trajectories, propose consequential next steps, and proactively co-direct inquiry as the shared exploration evolves. 
Moreover, extending human-AI collaboration beyond text-only chat interfaces to include other modalities fundamentally changes how humans perceive and engage with information.
In such mixed-modality interactions (\eg, text and direct manipulation), understanding no longer arises from passive interpretation but from active exploration, where users iteratively manipulate visual structures, pose follow-up questions, and refine their mental models in response to emerging patterns.
This exploratory mode couples perception and reasoning, enabling insights to be constructed through user interactions rather than merely received, and creating conditions for deeper understanding and the discovery of new knowledge.

Although evaluated here using science of science data, the principles underlying LLM-native figures are domain-agnostic. 
Any scientific field that produces structured data and visual representations can adopt this framework to unify computation, visualization, and reasoning. 
Extensions could include integration with simulation workflows, uncertainty quantification, and cross-domain data fusion.

Technically, future work may focus on (1) scalability, enabling artifact storage and versioning at large analytical scales; (2) semantic generalization, where LLMs learn to reason over increasingly complex visualization grammars and data modalities; and (3) collaborative reasoning, where multiple human or AI teammates interact over shared artifact graphs. 
Each direction expands the scope of computational transparency from individual analyses to entire ecosystems of scientific collaboration.

Several limitations warrant consideration. 
First, the current implementation relies on large language models that may generate erroneous code or misinterpret ambiguous queries; while guardrails mitigate these risks, full reliability remains an open challenge. 
Second, the framework presumes structured data and declarative visualization grammars, limiting immediate applicability to unstructured or domain-specific data types such as images or molecular structures. 
Finally, our evaluation focuses on exploratory tasks; formal assessment in confirmatory scientific studies will be essential to validate its broader impact. 
\wyf{Addressing these challenges will require collaboration across communities in AI, visualization, and computational science to realize the broader value of LLM-native figures: establishing a common infrastructure for reasoning in which transparency and provenance are built into the artifacts of discovery themselves.}





\section{Methods} 
\label{section:methods}
%


The system consists of three modules: the hybrid user interface, the multi-agent LLM engine, and the data management module (Fig.~\ref{fig:Figure8}). 
We describe each in more detail below.

\subsection{Hybrid User Interface}
\label{conceptual-model-hybrid-ui} 
We design a hybrid interface that couples an LUI and a GUI (Fig.~\ref{fig:Figure8}b), serving as the system frontend through which users interact directly with the system.

\textbf{Language User Interface (LUI)} (Fig.~\ref{fig:Figure8}b-left). 
By allowing humans to ask open-ended questions in natural language, the LUI lowers the cognitive and technical barriers to data exploration. 
In addition to traditional chatbot panels, the interface includes inline conversational widgets, which allow dialogue to unfold within the visual context itself.
Users can initiate queries via chat, request AI-generated analytical suggestions, or interact directly with visualizations to ask follow-up questions, request AI recommendations, and coordinate and filter across multiple visualizations. 
In this way, conversation becomes an analytical instrument rather than a passive query mechanism. 

\textbf{Graphical User Interface (GUI)} (Fig.~\ref{fig:Figure8}b-right).  
While language captures ideas, visualization externalizes them by transforming abstract data into perceptible structures, providing an intuitive and efficient medium for uncovering patterns that may be difficult to articulate through statistics alone~\cite{anscombe1973graphs}. 
The GUI enables users to interact directly with LLM-native figures through standard visualization operations such as brushing, filtering, clicking, and zooming. 
Through these interactions, users can quickly and accurately access underlying data through visual elements (\eg, dots in a scatterplot and bars in a bar chart)~\cite{shneiderman1983direct, munzner2014visualization}. 
LLM-native figures thus enable visual data exploration while tracing the associated analytical procedures and data. 
%
In addition to individual figures, the system supports data-driven artifacts (Sec.~\ref{subsection:conceptual-framework}-Data-Driven Artifact and Provenance), which are stored as linked figures with associated coordination rules and interaction histories.

\begin{figure}[!t]
    \centering
    \includegraphics[width=1\linewidth]{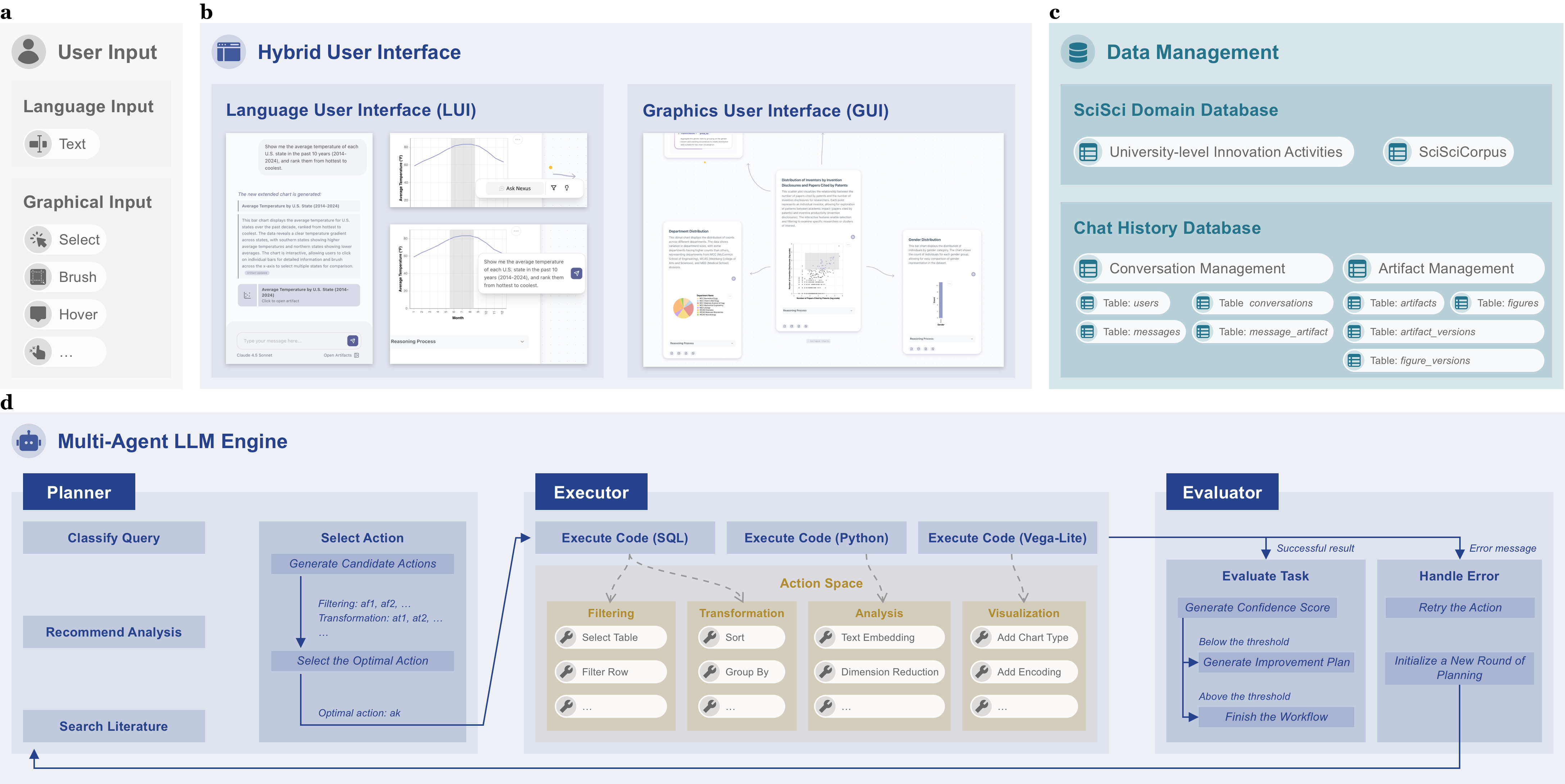}
    \caption{
    \bodyfigurelabel{fig:Figure8}
    \textbf{System design in \systemname.}
    \textbf{(a)} User input types include natural language input and graphical input. 
    \textbf{(b)} The Hybrid User Interface Module serves as the frontend of the system.  
    Left: natural language interaction using a traditional chatbot and an inline chatbot. 
    Right: graphical interaction using interactive dashboards (\eg, click, brush, and hover).
   \textbf{(c)} The Data Management Module manages SciSci domain datasets and knowledge and chat history.
    \textbf{(d)} The Multi-Agent LLM Engine serves as the backend of the system, including three agents and a set of tools. 
    }
\end{figure}

\subsection{Multi-Agent LLM Engine} 
\label{subsection:method-llm}
Creating this hybrid interface requires an AI infrastructure capable of understanding heterogeneous user inputs, reasoning over complex data, and generating data insights in real time. 
Thus, we design a multi-agent LLM engine that coordinates the communication between the user and the large-scale underlying data, leveraging recent advances in LLM-based reasoning~\cite{yao2023react}, planning~\cite{yao2023tree}, tool use~\cite{schick2023toolformer}, and self-improvement~\cite{madaan2023self, shinn2023reflexion} to support a wide range of analytical intents and exploration pathways. 
The engine includes three core agents, \planner, \executor, and \evaluator, which operate at an \actionname{action}level of granularity (Fig.~\ref{fig:Figure8}d). 
Together, they form an automated reasoning loop that spans data analysis, visual insight generation and manipulation, dashboard coordination, and iterative exploration.

\textbf{Multi-Agent Workflow.}
The \planner Agent serves as the direct intermediary between the user interface and the analytical backend. 
It infers a user's intentions by taking the user's natural language queries and direct manipulation inputs (\eg, brushing and filtering) and formulating an initial execution strategy. 
Specifically, it performs query triage across three distinct pathways.
For high-level complex analytical goals, it decomposes the query into structured sub-questions and returns them to users for clarification or confirmation. 
For exploratory scenarios where users seek guidance on subsequent analytical steps, it recommends potential next steps based on exploration history and intermediate results. 
For low-level questions, it initiates action planning, selection, and execution procedures. 

To accomplish these tasks, the agent leverages four specialized tools: 
(1) the \toolname{user_query_classifier()} tool classifies user queries into high-level (requiring decomposition) and low-level (directly executable) categories, enriched by domain knowledge retrieved via literature search; 
(2) the \toolname{ai_recommender()} tool proposes follow-up exploration suggestions based on user intent, multimodal data context, and domain knowledge; 
(3) the \toolname{literature_search()} tool~\cite{shao2025sciscigpt} retrieves and summarizes relevant literature from the SciSci domain using retrieval-augmented generation (RAG), thereby grounding the planning process in domain-specific context; 
(4) the \toolname{action_selector()} tool utilizes a tree-based planning strategy that synthesizes principles from tree-of-thoughts reasoning and search~\cite{yao2023tree, zhuang2023toolchain}. 
The tool constructs a dynamic search tree $\mathcal{T}$, where each node represents an atomic analytical action along with its execution results, and edges encode temporal dependencies between actions. 
This design enables exploration of multiple reasoning pathways by navigating a structured and comprehensive decision space (Supplementary Note 1.2). 

The \executor Agent executes the selected action by invoking the corresponding tool to generate the appropriate code (\textit{SQL} for data filtering and transformation, \textit{Python} for analysis and modeling, and \textit{Vega-Lite} for data visualization~\cite{satyanarayan2016vega}).
These code snippets are then executed within three isolated sandboxes, \toolname{execute_SQL()}, \toolname{execute_Python()}, and \toolname{execute_VegaLite()}, to produce intermediate outputs in LLM-native figures, including dataframes, visualization specifications, rendered images, and textual insights, among others. 

The \evaluator Agent incorporates the principle of self-reflection~\cite{shinn2023reflexion} from agent design frameworks to evaluate the outcomes of executed actions and guide subsequent planning. 
It handles two types of action outcomes: successful and failed executions. 
When an action is executed successfully, the agent assesses how well the result answers the user’s query. 
This assessment is made by the LLM which produces a confidence score between 0 and 1, along with an explanation and any remaining plans. 
If the score exceeds a predefined threshold, the action selection loop terminates. 
Otherwise, the execution result, evaluation score, rationale, and remaining plans are appended to the data context (Sec.~\ref{subsection:method-implementation}) and passed back to the \planner Agent for the next round of action selection.
%
If an action fails, the agent processes the error message and determines the next step, either retrying the same action in the \executor Agent, or initiating a new round of planning in the \planner Agent (\ie, finding an alternative solution path in the search tree $\mathcal{T}$). 
%
Together, the three agents coordinate to generate a sequence of actions that answer user queries, enhanced by advanced prompt design and context management (Sec.~\ref{subsection:method-implementation}).

\textbf{Action Space.}
We define a compositional action space consisting of atomic operations that serve as the fundamental execution primitives for data-driven exploration tasks. 
These actions are dynamically composed, sequenced, and executed by the multi-agent workflow to support diverse analytical tasks, including data filtering, transformation, modeling, visualization, and cross-visualization coordination. The space is designed to be refinable and extensible for adaptation across different domains~\cite{gao2025democratizing}. 
This action-level design improves reasoning accuracy and execution stability by constraining the solution space to well-defined operations amenable to optimization-based planning in the \planner Agent, while simultaneously enhancing algorithmic transparency, enabling users to inspect and validate each analytical step~\cite{zhuang2023toolchain, hong2024data, chen2021vizlinter}.
%
Specifically, we design four types of actions derived from general data science tasks: \textit{data filtering}, \textit{data transformation}, \textit{data analysis}, and \textit{data visualization}, building on techniques such as NL2SQL~\cite{liu2024survey, hong2024next}, AI4VIS~\cite{wu2021ai4vis}, and LLM for data science~\cite{guo2024ds, hong2024data} (Supplementary Note 1.1). 
For \textit{data visualization}, we include actions for selecting the chart type (\actionname{add_chart_type}) and interaction type (\actionname{add_params}), as well as for attaching or updating data (\actionname{add_data} and \actionname{update_data}) and visual encodings (\actionname{add_encoding} and \actionname{update_encoding}). 
%
Building on these atomic actions, the LLM engine can easily interpret the user's analytical intent using natural language instructions and graphical interactions, and use bidirectional mapping (Sec.~\ref{subsection:conceptual-framework}-Bidirectional Mapping) to generate, refine, and extend figures during the exploration process.

\textbf{Coordination Across Figures.} 
The system supports the generation and coordination of multiple linked figures within a single data-driven artifact, enabling coherent multi-view exploration. 
Coordination is triggered in two stages during visualization-driven analysis.
First, when users generate a new figure by interacting with an existing one, the system jointly interprets GUI interactions and natural-language instructions to infer analysis intent. 
These interaction-derived constraints are injected into subsequent analytical actions, ensuring that newly generated figures remain semantically aligned with prior exploration. 
To support iterative analysis, the LLM engine can flexibly query both the underlying SciSci domain database and a temporary workspace derived from prior figures (Fig.~\ref{fig:Figure1}c–d).
Second, to enable persistent cross-figure coordination, the system records a coordination schema for each pair of linked figures at generation time, derived from the executed action sequence. 
When users later update GUI selections in an initiating figure, the system retrieves the corresponding schema and re-executes the stored analytical workflow, which includes both data filtering and downstream analysis steps, to regenerate all coordinated figures with consistent, up-to-date results. 
This design supports flexible cross-figure interaction while preserving provenance and reproducibility of the analytical process (Supplementary Note 1.3).

\subsection{Data Management}
\label{subsection:method-data-module}
The Data Management Module in \systemname rests on three interlinked databases (Fig.~\ref{fig:Figure8}c).
The first two databases are domain-specific, while the third is generalizable across domains.
(1) The SciSci relational database is a domain-specific database that provides the empirical substrate for analysis. 
It is a university-level dataset capturing innovation activities and researcher information~\cite{wang2023innovationinsights}. 
\newlyadded{The data are collected and preprocessed from multiple sources, including public datasets such as SciSciNet~\cite{lin2023sciscinet}, Microsoft Academic Graph (MAG)~\cite{wang2019review}, PatentsView~\cite{patentsview}, Reliance on Science~\cite{marx2020reliance, marx2022reliance}, as well as proprietary institutional datasets and subscription-based commercial databases (Supplementary Note 2).}
%
(2) SciSciCorpus~\cite{shao2025sciscigpt} is a vector database for SciSci literature, in which full-text papers are chunked, indexed, and embedded to support retrieval-augmented generation (RAG). This enables more accurate and context-aware reasoning by LLM agents when addressing complex analytical queries in the SciSci domain.  
%
(3) A Chat History Database is tailored to support LLM-native figures and data-driven artifacts for non-linear exploration processes, capturing the evolving logic of discovery (Supplementary Note 1.5). 
It includes a \code{conversations} table that stores all historical sessions created by the user, and a \code{messages} table that stores individual question-answer pairs for each user session, establishing the conversational foundation. 
Each message links to corresponding artifact entries in three artifact-related tables (\code{message_artifact}, \code{artifacts}, and \code{artifact_versions}), which maintain comprehensive metadata for each data-driven artifact, including a list of figure identifiers, coordination relationships, and schemas that define cross-figure interactions. 
The figure identifiers in each artifact link to two tables, \code{figures} and \code{figure_versions}, which store the components of each LLM-native figure: visualization, code, dataset, and meta information (Sec.~\ref{subsection:conceptual-framework}), as well as action-level reasoning, code, and results. 
It also includes two tables, \code{artifact_versions} and \code{figure_versions}, that
record the exploration history as a list of temporal snapshots of artifact-level and figure-level states, such that each user input generates a new record, preserving the complete exploration history. 
This architecture enables complete reconstruction of analytical pathways, allowing users to trace their discovery processes and revert to previous analytical states transparently.

\subsection{Implementation Details} 
\label{subsection:method-implementation} 
\textbf{Technology Stack.}
The system is implemented as a full-stack web application utilizing modern development frameworks and cloud infrastructure. 
The frontend employs React.js~\cite{React} for component-based user interface development, Redux.js~\cite{Redux} for state management, and Vega-Lite for declarative visualization rendering~\cite{satyanarayan2016vega}. 
The backend architecture is built on the Flask framework~\cite{Flask} with Python~\cite{Python}, providing RESTful API services and data processing capabilities. 
Specifically, the multi-agent workflow leverages LangChain~\cite{LangChain} and LangGraph~\cite{LangGraph} for complex agent coordination, with Claude Sonnet 4.5~\cite{AnthropicAPI} serving as the primary language model (Supplementary Note 1.6). 
The system architecture supports model substitution, enabling integration with alternative LLM providers as needed. 
Data storage employs a hybrid approach: Google BigQuery~\cite{BigQuery} manages relational SciSci domain datasets and user chat histories, while Pinecone~\cite{Pinecone} serves as the vector database for the retrieval-augmented generation module, storing chunked domain literature for semantic search and knowledge retrieval.

\textbf{Action-Based Streaming Output.} 
To balance analytical transparency with user experience, we implement action-level streaming output that selectively displays critical workflow information without overwhelming users with lengthy LLM reasoning. 
The interface presents only essential decision points, specifically action selection and action sequence generation processes, providing users with meaningful insights into analytical behavior.
Upon completion of each user query, the system delivers comprehensive results including SQL queries, Python code, processed dataframes, visualization images, and structured Vega-Lite JSON outputs. 
This approach ensures users can monitor system progress through high-level action updates while accessing detailed technical artifacts for validation and reuse. 

\textbf{Prompt and Context Design.}
We employ four prompt engineering techniques to optimize our multi-agent workflow performance~\cite{white2023prompt, sahoo2024systematic}: structured formatting, chain-of-thought reasoning, few-shot learning, and dynamic prompting. 
In addition, to make the context compact, we developed a structured data context that stores only essential information about each executed action, including action indices, objectives, execution status, results, evaluation scores and rationale, and remaining tasks (Supplementary Note 1.4). 


\newpage

\section*{Data Availability}
Data necessary to reproduce all plots will be made freely available. 

\section*{Code Availability}
Code necessary to reproduce all plots will be made freely available.

\section*{Acknowledgments}
We thank all members of the Center for Science of Science and Innovation (CSSI) at Northwestern University for helpful discussions, and Alyse Freilich for her careful editing and valuable feedback. 
This work is partly supported by the National Science Foundation (award number 2404035; D.W.). 
Any opinions, findings, and conclusions or recommendations expressed in this material are those of the author(s) and do not necessarily reflect the views of the funders. 

\section*{Author Contributions}
Y.W., R.S., E.S., Y.Q., and H.L. designed the methodology and conducted the investigation. 
Y.W., R.S., and E.S. developed the system. 
D.W. and N.C. conceived the study. 
D.W. administered the project. 
All authors contributed to writing, reviewed the manuscript critically for important intellectual content, and approved the final version for publication. 

\section*{Competing Interests}
The authors declare no competing interests. 



\vspace{4pt}
\noindent{\bfseries References}\setlength{\parskip}{12pt}%

\bibliography{mybib}



\beginedfigures


\end{document}